\documentclass[twocolumn,aps,prc,superscriptaddress,reprint]{revtex4-1}
\makeatletter

\newcommand{\Rmnum}[1]{\expandafter\@slowromancap\romannumeral #1@}
\makeatother
\usepackage{blindtext}
\usepackage{graphicx}
\usepackage{dcolumn}
\usepackage{bm}
\usepackage{booktabs}
\usepackage{color}
\usepackage{mathrsfs}
\usepackage{ulem}
\usepackage{verbatim}
\usepackage{indentfirst}
\usepackage{float}
\usepackage[FIGTOPCAP]{subfigure}
\usepackage[colorlinks,
           bookmarks=true,
           linkcolor=blue,
           urlcolor=blue, 
            anchorcolor=black,
            citecolor=blue
            ]{hyperref}
\usepackage{hypcap}
\usepackage{CJK}
\usepackage{amsmath}

\renewcommand\sout{\bgroup \color{red} \ULdepth=-.5ex \ULset}

\newcommand{\commentout}[1]{}

\usepackage{epsfig}
\graphicspath{{plots/}}

\begin{document}
\title{Beam energy dependence of the squeeze-out
effect on the directed and elliptic flow
in Au+Au collisions in the high baryon density region}

\author{Chao Zhang}
\affiliation{ Institute of Particle Physics and Key Laboratory of Quark\&Lepton Physics (MOE),Central China Normal University, Wuhan 430079, China}
\author{Jiamin Chen}
\affiliation{ Institute of Particle Physics and Key Laboratory of Quark\&Lepton Physics (MOE),Central China Normal University, Wuhan 430079, China}
\author{Xiaofeng Luo}
\email{xfluo@mail.ccnu.edu.cn}
\affiliation{ Institute of Particle Physics and Key Laboratory of Quark\&Lepton Physics (MOE),Central China Normal University, Wuhan 430079, China}
\author{Feng Liu}
\affiliation{ Institute of Particle Physics and Key Laboratory of Quark\&Lepton Physics (MOE),Central China Normal University, Wuhan 430079, China}
\author{Yasushi Nara}
\affiliation{Akita International University, Yuwa, Akita-city 010-1292, Japan}
\affiliation{Frankfurt Institute for Advanced Studies, 
D-60438 Frankfurt am Main, Germany}

\begin{abstract}
We present a detailed analysis of
the beam energy dependence of the mechanisms for the generation of
directed and elliptic flows
in Au+Au collisions 
focusing on the role of hadronic rescattering and spectator
shadowing
within a microscopic transport model JAM
with different equation of state.
A systematic study of the beam energy dependence 
is performed for Au+Au collisions
at $\sqrt{s_{NN}} =2.3 - 62.4$ GeV.
The transition of the dynamical origin of the directed flow is observed.
We find that the initial Glauber type nucleon-nucleon collisions
generate negative $v_1$ for nucleons at midrapidity  due to the presence of
spectator matter, and this negative nucleon $v_1$ is turned 
to be positive by the meson-baryon interactions
at the beam energy region of $\sqrt{s_{NN}} < 30$ GeV.
In contrast, above 30 GeV there is no spectator shadowing
at midrapidity, and initial nucleon-nucleon collisions do not generate
directed flow, but subsequent rescatterings among produced particles
generate negative $v_1$ for nucleons.
It is demonstrated that
negative pion-directed flows are mostly generated by the interaction
with the spectator matter.
It is also shown that the squeeze-out effect is largely suppressed in the case
of softening, which leads to the enhancement of elliptic flow
around $\sqrt{s_{NN}}=5-7$ GeV.
The elliptic flow at midrapidity above 10 GeV is not influenced by
the squeeze-out due to spectator matter, 
while its effect is seen
at the forward rapidity range of $y/y_\mathrm{c.m.}>0.5$,
which decreases as beam energy increases.
\end{abstract}
\maketitle

\section{Introduction}
A new form of strongly interacting dense QCD matter
called Quark-Gluon plasma (QGP) is created
in the experiments in high-energy heavy ion collisions
at the Relativistic Heavy Ion Collider (RHIC)
and Large hadron Collider (LHC)~\cite{Arsene2005,Heinz:2013th,Busza:2018rrf}.
The lattice QCD calculations confirmed that at zero baryon density,
the transition from hadronic matter to QGP is a crossover~\cite{Aoki:2006we}.
The next challenge is to explore the phase diagram of QCD matter extending
to finite baryon density regions
by creating compressed baryonic matter (CBM)~\cite{Friman:2011zz}.
At nonvanishing baryon densities, a first-order and/or second-order
QCD phase transition together with a critical end point
have been speculated by many theoretical calculations~\cite{DHRischke2004}.
A first-order phase transition implies the existence of a strong softest point
in the equation of state (EoS)~\cite{CMHungPRL,DHRischke1995},
and it is conjectured that this softening effects may be seen in observables.
To find the evidence of a phase transition
and the presence of a critical end point,
various observables have been measured such as
particle ratios, moments of the conserved charges,
and collective flows~\cite{Gupta:2011wh,LKumar2013,Luo:2017faz}.
New CBM experiments are planned such as Beam Energy Scan
(BES \uppercase\expandafter{\romannumeral2} )
at RHIC~\cite{GOdyniecEPJ2015}, 
Facility for Antiproton and Ion Research (FAIR)
~\cite{CHhneJPCS2013,Ablyazimov:2017guv},
Japan Proton Accelerator Research Complex for Heavy Ion (J-PARC-HI)
~\cite{HSakoNPA2016}, 
SPS Heavy Ion and Neutrino Experiment (NA61-SHINE) at the Super-Proton Synchrotron (SPS)
~\cite{Turko:2018kvt},
and Nuclotron-based Ion Collider fAciilty (NICA)~\cite{VKekelidzeNPA2016}
 to explore the phase diagram at high baryon density region.

In this paper, we focus on anisotropic collective flows.
An analysis of the anisotropic collective flows~\cite{JHofmannPRL,HStoecker}
in non-central nuclear collisions appears to be
one of the most popular methods
in studying the properties of the hot and dense matter
since they are sensitive to the EoS in the early stages of nuclear
collisions~\cite{HStoecker2005,HGBaumgardt1975},
and thus considered to be a good probe to explore the properties of the QCD
matters.
The anisotropic flows are defined by the Fourier coefficients
of the expansion of the azimuthal distribution of particles
measured with respect to the reaction plane, 
which is spanned by the vector of the impact parameter and the beam direction
~\cite{SVoloshinZPC1996,Poskanzer58,Voloshin2008}:
\begin{align}\label{eq:definition1}
E\frac{d^{3}N}{d^{3}p}&=\frac{1}{2\pi}\frac{d^{2}N}{p_{T}dp_{T}dy}
  \left[1 
  +2\sum_{n=1}^\infty
 v_{n}\cos n[(\phi-\Psi_{RP})]\right]\,,
\end{align}
where $\phi$ is the azimuthal angle,
and $\Psi_{RP}$ is the reaction plane azimuthal angle.
The flow coefficients $v_{n}=\langle \cos n[(\phi-\Psi_{RP})]\rangle$
characterize the event anisotropy.
The directed flow parameter is defined as the first Fourier coefficient $v_{1}$ of the particle momentum distribution, 
and the second coefficient $v_{2}$ is referred to as elliptic flow.
  
Directed flow is very sensitive to the early dynamics of the heavy ion 
collisions~\cite{Voloshin2008}.
The excitation function of the nucleon directed flow at midrapidity
is predicted to have a minimum at a certain collision energy
in the hydrodynamical calculations
with a first-order phase transition  (1OPT)
~\cite{DHRischke1995, JBrachmannPRC2000, YBIvanovEPJ2003}.
Furthermore, when the system passes through the softest point of the EoS, 
the slope of the directed flow for nucleons turns to be negative
~\cite{JBrachmannPRC2000, JBrachmannEPJ2000, LPCsernaiPLB1999},
which is called the third flow~\cite{LPCsernaiPLB1999, LPCsernaiAPHA2005} 
or the anti-flow of nucleon~\cite{JBrachmannPRC2000,JBrachmannEPJ2000}.
This does not happen for the crossover transition~\cite{JSteinheimerPRC2014}.
Hence, it is predicted that the collapse of the directed flow 
is a signature of a QCD first-order phase transition.
The negative slope of protons in a first-order phase transition
has been also confirmed  within a microscopic transport model JAM,
in which the effects of EoS is incorporated by changing a
scattering style in the two-body collisions
~\cite{YNaraPRC2016,Nara:2016hbg}.
Recent measurements by
the STAR Collaboration~\cite{LAdamczykPRL2014, PShanmuganathanarxiv}
show negative proton $v_1$ at above $\sqrt{s_{NN}}\approx 10$ GeV
in the energy range of RHIC-BES program.
We note that in the microscopic transport models
RQMD~\cite{RJMSnellingsPRL2000}, UrQMD~\cite{MBleicherPLB2002}, 
and PHSD/HSD~\cite{VPKonchakovskiPRC2014}, 
the negative slope of proton ${v}_1$ at midrapidity 
is found at high bombarding energies $\sqrt{s_{NN}}\geq 27$ GeV,
which is caused by the certain amount of
degree of rapidity loss of incoming nucleons and positive space momentum
correlation~\cite{RJMSnellingsPRL2000}.
At the bombarding energies of $\sqrt{s_{NN}}$ $\leq$ 20 GeV
~\cite{VPKonchakovskiPRC2014, HPetersenPRC2006},
such microscopic transport models do not show a negative slope for nucleons.
Therefore, the negative proton $v_1$ slope at midrapidity
at $\sqrt{s_{NN}}\leq 20$ GeV can be only produced by theoretical models
which incorporate the effect of a first-order phase transition.

The elliptic flow also provides information about the early stages of
the collision~\cite{Sorge78,Teaney86,Heinz0907,HSorgePRL1999}, 
and it is one of the most extensively studied observables
in relativistic nucleus-nucleus collisions.
At lower beam energies of $E_\text{lab} \leq 5$ AGeV,
shadowing effects by the spectator matter have been known
to play a essential role for directed and elliptic flows.
It is known that reflection of pion by the nucleon is the dominant
origin of the negative directed flow for pions~\cite{Bass:1993em}
at $E_\text{lab}\approx 1-2$ AGeV.
The presence of spectator matter is the origin of the squeeze-out
(out-of-plane emission)~\cite{Bass:1993ce,Hartnack:1994ce},
and elliptic flow can be even negative at lower energies.
In the high baryon density region, such as Alternating Gradient Synchrotron (AGS) energies,
the final strength of the elliptic flow is determined by the 
interplay between squeeze-out effect and in-plane emission~\cite{Sorge78}.
Enhancement of elliptic flow due to the softening of EoS
is predicted~\cite{Chen:2017cjw,Nara:2017qcg}, because softening of EoS
suppresses the squeeze-out effect.

So far a systematic study of the role of spectator matter
for a wide range of beam energies has not been performed.
In this paper, we shall study in detail the role of spectator shadowing on both
the directed and elliptic flows, together with its EoS dependence.
For this purpose, we utilize a microscopic transport model JAM 
to systematically study the collision dynamics emphasis on
the effects of EoS,
hadron rescattering, hadronic mean-field,
and interaction with spectator
to the anisotropic collective flows
in Au+Au collision at $\sqrt{s_{NN}}=2.3-62.4$ GeV.
We shall show that the shadowing effect by the spectator
 still plays an important role
for the generation of anisotropic flows at RHIC-BES energies of
$\sqrt{s_{NN}}<30$ GeV.
We investigate such effects by disabling meson-baryon ($MB$),
and meson-meson ($MM$) collisions, as well as the interactions between
participants and spectator matter.

The paper is organized as follow, 
after the brief description of the JAM model in section~\ref{sec:jam},
we compute the rapidity and transverse momentum distributions
of identified particles with different EoS in sec.~\ref{sec:v0}.
In Sec.~\ref{sec:spec5}, hadronic rescattering and spectator effects
together with EoS dependence are discussed for Au+Au collisions
at $\sqrt{s_{NN}}=5$ GeV.
Then we investigate the beam dependence of the spectator shadowing
on the directed and elliptic flows in Sec.~\ref{sec:beam}.
Finally, a summary will be given in section~\ref{sec:summary}.

\section{JAM model}
\label{sec:jam}

We employ a hadronic transport model JAM~\cite{YNaraPRC2000} 
that is developed to simulate relativistic nuclear collisions from 
initial stage to final state interaction in hadronic gas stage. 
Similar to other transport models~\cite{HSorgePLB1997,SABassNP1998, 
MBleicherJPG1999, ZWLinPRC2005, WCassingNPA2009}, 
the particle production in JAM is modeled by resonance and string production 
and their decay, and the particles including produced ones
can interact with each other by the two-body collisions
~\cite{THiranoPTEP2012}.

The effects of the equation of state
have been implemented by two different approaches:
the nuclear mean-field and modified two-body scatterings.
The nuclear mean-field potential is implemented along the lines of
the simplified version~\cite{Maruyama:1996rn,Mancusi:2009zz} of
the relativistic quantum molecular dynamics approach~\cite{RQMD}.
In this approach, our Hamiltonian is given by the sum of
single-particle energy:
\begin{equation}
 H =\sum_{i=1}^{N}\sqrt{\bm{p}^2_i + m_i^2 + 2m_iV_i}
\end{equation}
and the following equations of motion 
\begin{align}
  \frac{d\bm{r}_i}{dt} &= \frac{\partial H}{\partial\bm{p}_i}
   =\frac{\bm{p}_i}{p^0_i}
   + \sum_{j=1}^N\frac{m_j}{p^0_j}
      \frac{\partial V_j}{\partial\bm{p}_j},\nonumber\\
  \frac{d\bm{p}_i}{dt} &= -\frac{\partial H}{\partial\bm{r}_i}
   = -\sum_{j=1}^N\frac{m_j}{p^0_j}
      \frac{\partial V_j}{\partial\bm{r}_j}.
\end{align}
are numerically solved.
The relative distances in the two-body center-of-mass frame
are used in the argument of the potentials $V_i$:
\begin{align}
 -q_{Tij}^2 &= -(q_{i}-q_j)^2
    + \frac{[(q_i-q_j)\cdot (p_i+p_j)]^2}{(p_i+p_j)^2},\\
 -p_{Tij}^2 &= -(p_{i}-p_j)^2
    + \frac{[(p_i-p_j)\cdot (p_i+p_j)]^2}{(p_i+p_j)^2},
\end{align}
where $q_i$ and $p_i$ are the four-vectors for the coordinate and momentum of the
$i$-particle, respectively.
For the potential $V_i$,
the Skyrme-type density dependent and Lorentzian-type momentum-dependent
mean-field potential~\cite{CGalePRC1987}
are implemented in the model~\cite{MIssePRC2005,YNaraNPA2016}.
In this work, we use the parameter set used in Ref.~\cite{YNaraNPA2016},
which yields the nuclear incompressibility of $K=272$ MeV.

As a second approach, we control the pressure of the system by 
changing the scattering style in the two-body collisions~\cite{HSorgePRL1999}. 
The pressure of the system with volume $V$
in which only two-body scatterings happen
can be estimated
by the Virial theorem~\cite{PDanielewiczPRC1996}
\begin{eqnarray}
P=P_f+\frac{1}{3TV} \sum_{(i,j)}\,
  (\bm{p}'_i-\bm{p}_i)\cdot (\bm{r}_i-\bm{r}_j)
\end{eqnarray}
during the time interval $T$,
where $(\bm{p}'_i-\bm{p}_i)$ is a momentum transfer,
and $(\bm{r}_i-\bm{r}_j)$ is a relative coordinate
between two colliding particles $i$ and $j$
in the center-of-mass frame.
$P_f$ is the pressure from the free streaming contribution.
Thus the repulsive orbit 
$(\bm{p}'_i-\bm{p}_i)\cdot (\bm{r}_i-\bm{r}_j) >0$ enhances the pressure,
while the attractive orbit
$(\bm{p}'_i-\bm{p}_i)\cdot (\bm{r}_i-\bm{r}_j) <0$
reduces the pressure.
In the standard transport approach, 
the azimuthal angle of the two-body scattering is randomly chosen.
Consequently,the pressure generated by this scattering
is zero in average which leads to the free hadron gas EoS.
This immediately implies that one can control the pressure
by appropriately choosing the scattering style.
The selection of the repulsive orbit in the two-body collision
~\cite{ECHalbertPRC1981, MGyulassyPLB1982, DEKahanaPRC1997}
can simulate the effect of repulsive potentials.
While it is shown that selecting attractive orbit for all two-body
scattering yields the compatible amount of softening of a EoS with
a first-order phase transition, thus it
mimics the effects of a first-order phase transition
~\cite{YNaraPRC2016}.
In this work, the ``attractive orbit" mode in JAM refers to the simulation
in which attractive orbits are selected for all two-body scatterings
without imposing any conditions.

The EoS of the system can be controlled by the formula
in Ref.~\cite{HSorgePRL1999} by the following constraints
in the two-body scattering
\begin{equation}
\Delta P = \frac{\rho}{3(\delta\tau_i + \delta\tau_j)}
 (\bm{p}'_i-\bm{p}_i)\cdot (\bm{r}_i-\bm{r}_j) \,,
  \label{eq:eosmod}
\end{equation}
where $\Delta P$ is the pressure difference from the free streaming pressure,
$\rho$ is the local particle density and $\delta\tau_i$
is the proper time interval of the $i$-particle
between successive collisions.
We show that a given EoS can be simulated
by choosing azimuthal angle according to
the constraint in ~Eq.(\ref{eq:eosmod})
in the two-body scattering process~\cite{Nara:2016hbg}.
The main advantage of this approach is to be able to simulate any given EoS
with a numerically efficient way as far as there are many two-body collisions,
which happens in heavy-ion collision such as Au+Au collisions.
We use the same EoS used in Ref.~\cite{Nara:2016hbg} to simulate
1OPT (JAM/1OPT) based on Eq.~(\ref{eq:eosmod}) in this paper.

\section{Effects of hadronic mean fields and softening}
\label{sec:v0}

In this section, we investigate the effects of EoS
on the spectra of identified particles by utilizing the mean-field simulation
and modified scattering style approach.
\begin{figure}[!htb]
\setcounter{figure}{1}  
     \subfigure[]{
    \includegraphics[scale=0.40]{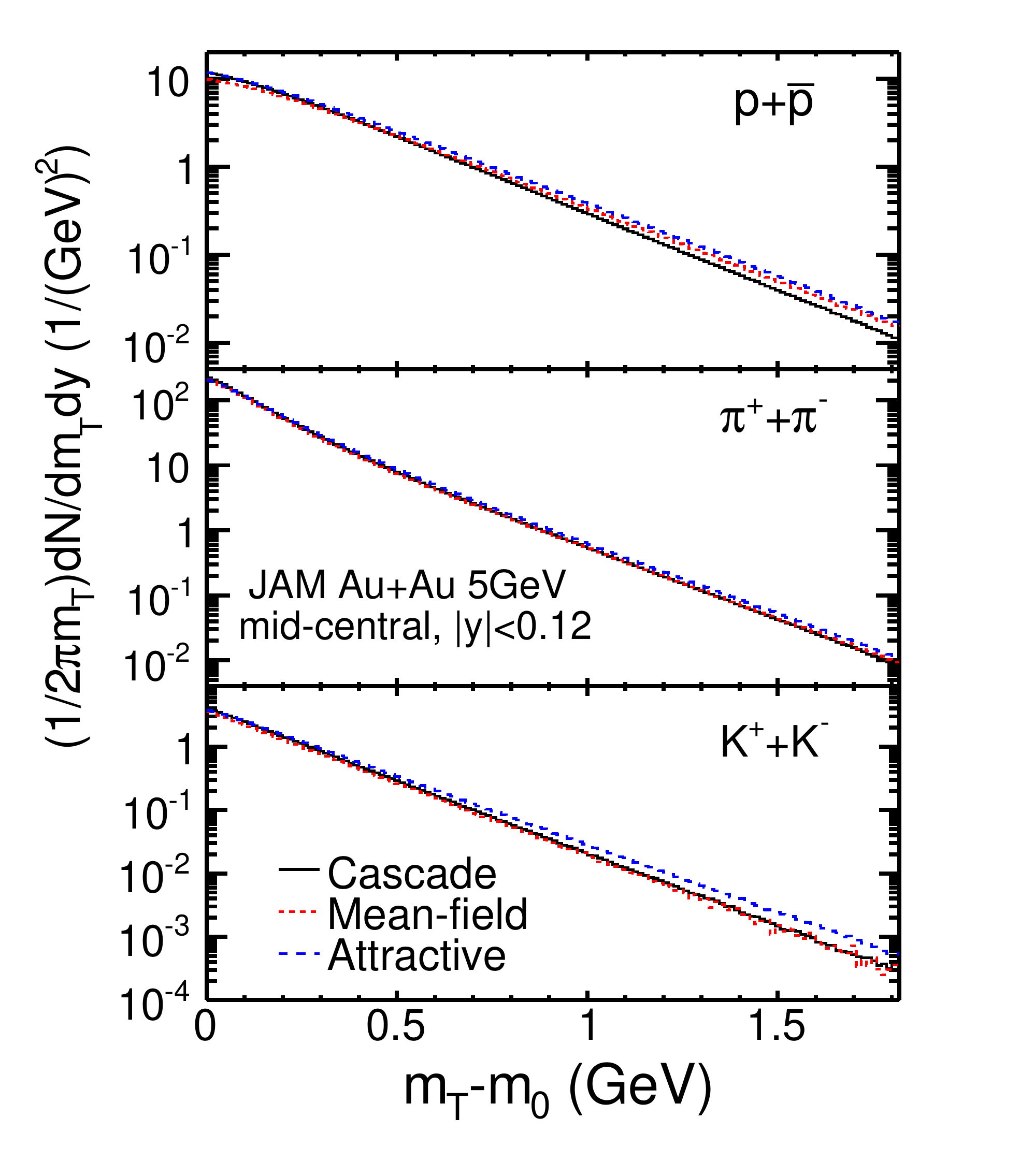}
    \label{fig:subfig:a}
    }
 \hspace{0.3in}
 \subfigure[]{
    \includegraphics[scale=0.40]{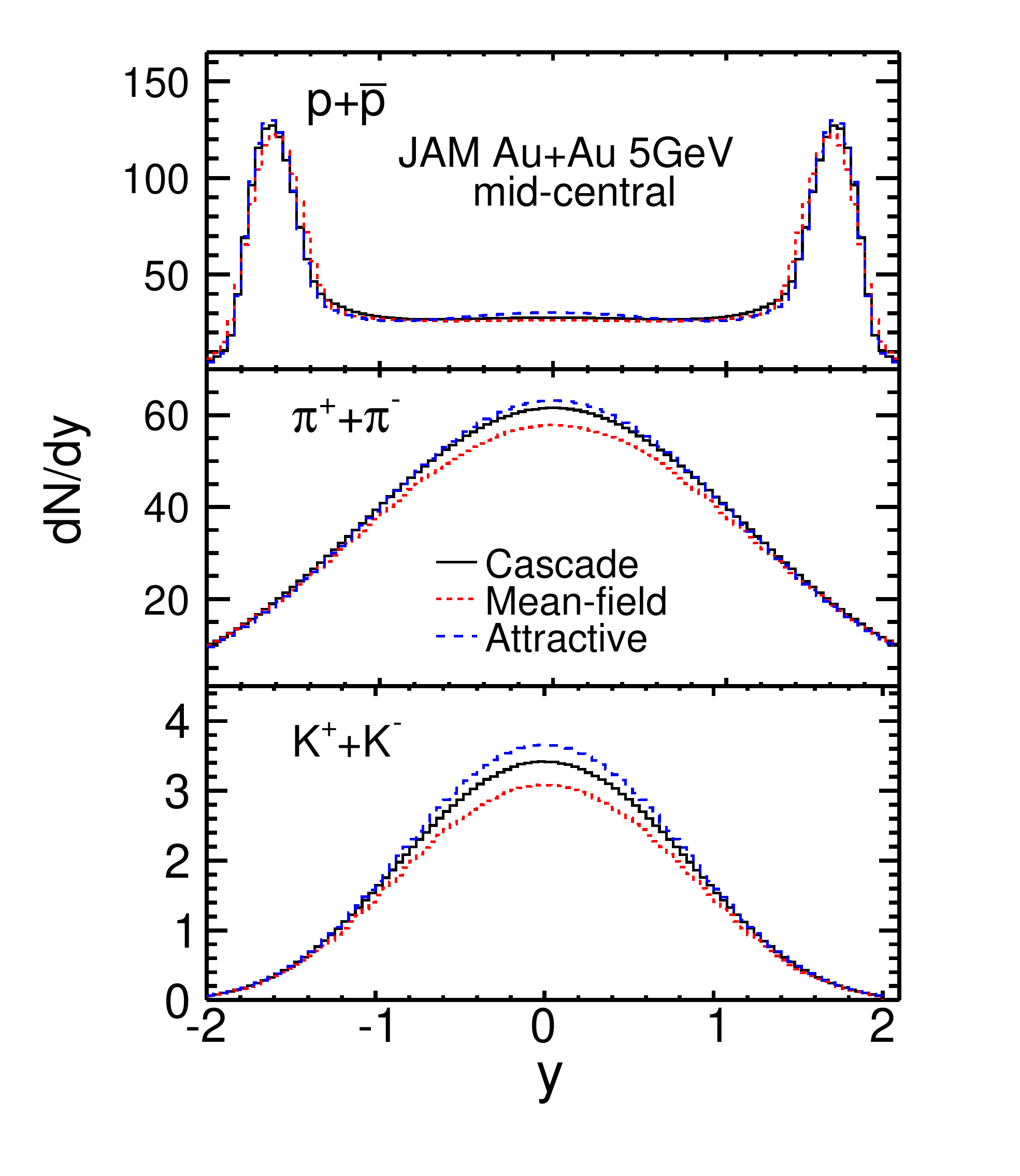}
    \label{fig:subfig:b}
   }
   \vspace{-0.2in}
\caption{(a) Transverse momentum spectrum at midrapidity ($|y|<0.12$),
and (b) rapidity distributions
of identified particles; proton, pion, kaon in midcentral (10--40\%) Au+Au collision
at $\sqrt{s_{NN}}$ = 5 GeV from JAM are compared with different EoS; cascade, with mean-field potential,
and cascade with attractive orbit mode.
}
\end{figure}


\subsection{EoS dependence}

We first compare the transverse mass spectrum
and the rapidity distribution of identified particles
including protons, pions, and kaons to see the effects 
of EoS on the gross dynamics of the
collision.
Figure~\ref{fig:subfig:a} shows the transverse mass distribution
at mid-rapidity $|y|<0.12$
for $p+\bar{p}$, $\pi^++\pi^-$ and $K^++K^-$ 
in midcentral (10-40\%) Au+Au collision at $\sqrt{s_{NN}}=5$ GeV
from JAM cascade, mean-field potential, and attractive orbit simulations.
The softening effect predicts the enhancement of collective transverse flow
for all particles.
This is understood by the longer interaction time due to the slow compression
and expansion of the system~\cite{Nara:2017qcg} by the softening.
Radial flow is generated from early to later stages of the collisions,
unlike anisotropic flows which are sensitive to the early pressure.
In addition, it is also proportional to the $pdV$ work, thus the radial flow
get larger as the system volume becomes larger.
The enhancement of the transverse flow
is also reported within hydrodyanmical approaches
with a first-order phase transition
~\cite{Petersen:2009mz,Ivanov:2013yla,Ivanov:2016sqy}.
JAM with hadronic mean field mode also predicts the harder slope
of protons, which is due to the repulsive potential.

Rapidity distributions of identified particles are displayed 
in Figure~\ref{fig:subfig:b}.
Softening effect is also seen in the rapidity distributions for all particles:
protons, pions, and kaons yield are enhanced at mid-rapidity,
while it is reduced at forward rapidity region
making total particle yields the same.
The effect of hadronic mean-field is opposite:
it reduces the particle yield at mid-rapidity, while it enhances
the yield at forward region.

\begin{figure}[!htb]
 \begin{center}
  \setcounter{figure}{1}  
   \includegraphics[scale=0.45]{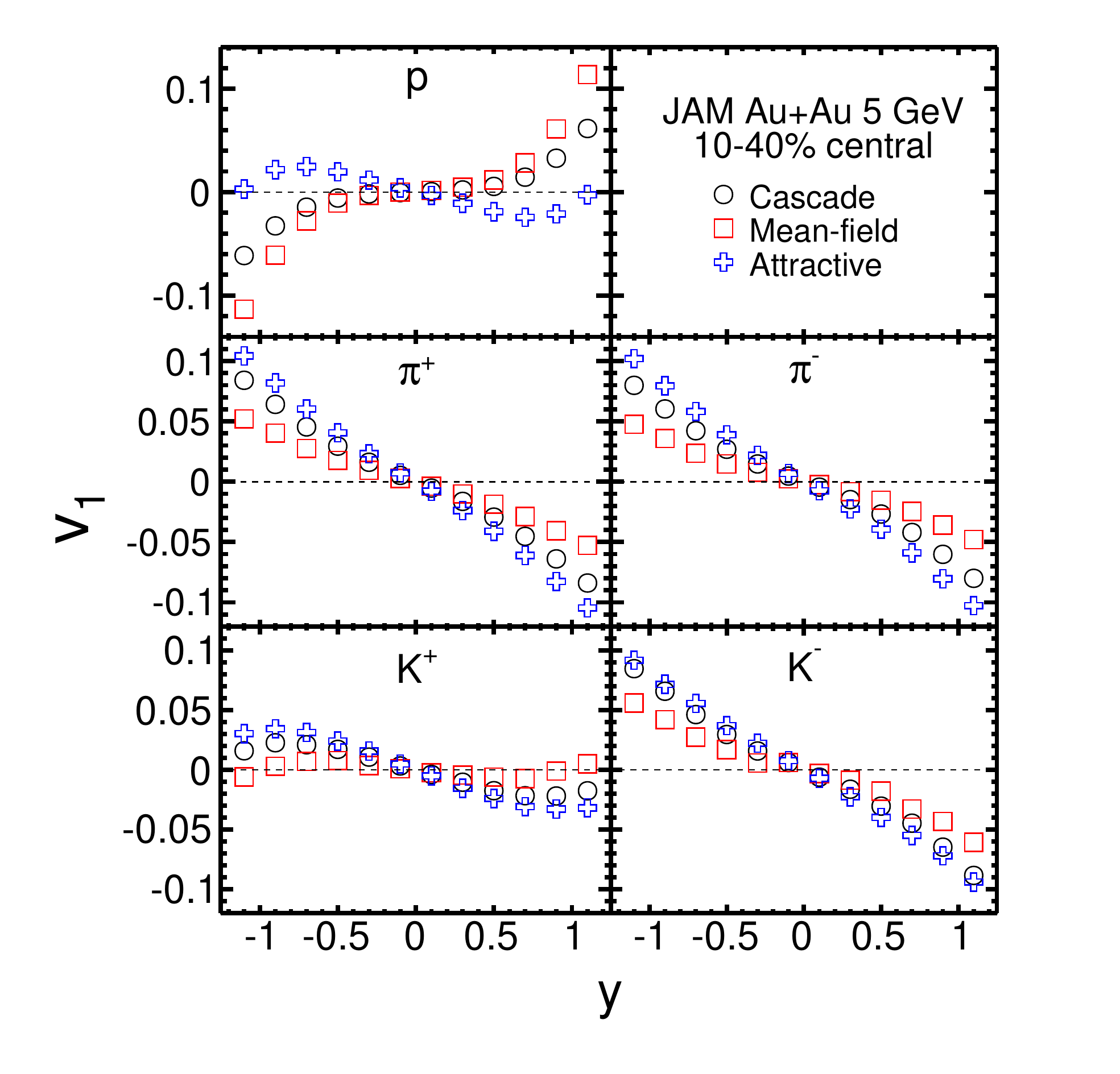}
   \caption{Directed flows as a function of rapidity
   in midcentral (10-40\%) Au+Au collision at $\sqrt{s_{NN}}$=5 GeV 
   from JAM cascade (circles), 
   JAM with mean-field potential (squires),
   and JAM cascade with attractive orbit (crosses).
   The left and right panels show the results for identified particles
   $p$, $\pi^{+}$, $K^{+}$ and antiparticles
   $\pi^{-}$, $K^{-}$ respectively.}
   \label{fig:directedflow}
  \end{center}
 \end{figure}

We see some effects of EoS on the transverse momentum
and rapidity distributions, but it is not a sizable effect.
However, the EoS effect is dramatic for the collective flows.
Sensitivities of the EoS on the elliptic flows of identified particles are
studied in Ref.~\cite{Chen:2017cjw}.

In Fig.~\ref{fig:directedflow},
rapidity dependence  of the directed flows
for identified particles ($p$, $\pi^{+}$, $K^{+}$) and 
corresponding antiparticles ($\pi^{-}$, $K^{-}$) 
in midcentral (10-40\%) Au+Au collision at $\sqrt{s_{NN}}=5$ GeV
are compared with different JAM modes:
cascade, mean-field potential,
and cascade with attractive orbit.
It is seen that the attractive orbit simulation
predicts a negative slope of protons that is
consistent with the results from a first-order phase transition
in Ref.~\cite{Nara:2017qcg}.
However, 
the mean-field potential enhances the positive $v_1$ slope for protons
due to the repulsive potential,
whereas, it reduces the slope for pions and kaons.
There is a weak EoS dependence on the kaons and pions $v_1$ that is
always negative slope.
The difference between the kaon and antikaon $v_1$ is due to the difference
of the cross section: the antikaon can form resonances with nucleons
similarly  to pion-nucleon scatterings,
thus it has larger cross sections than kaon-nucleon collisions.
This also explains the similarity of the antikaon $v_1$ to 
the pion $v_1$. 
We note that the small kaon-nucleon cross section
produces smaller $v_1$ for positive kaons compared to protons and pions.

\subsection{Comparison to the data}

In this section, we compare our results on the flows from different modes
with experimental data.

\begin{figure}[th]
 \begin{center}
   \includegraphics[scale=0.5]{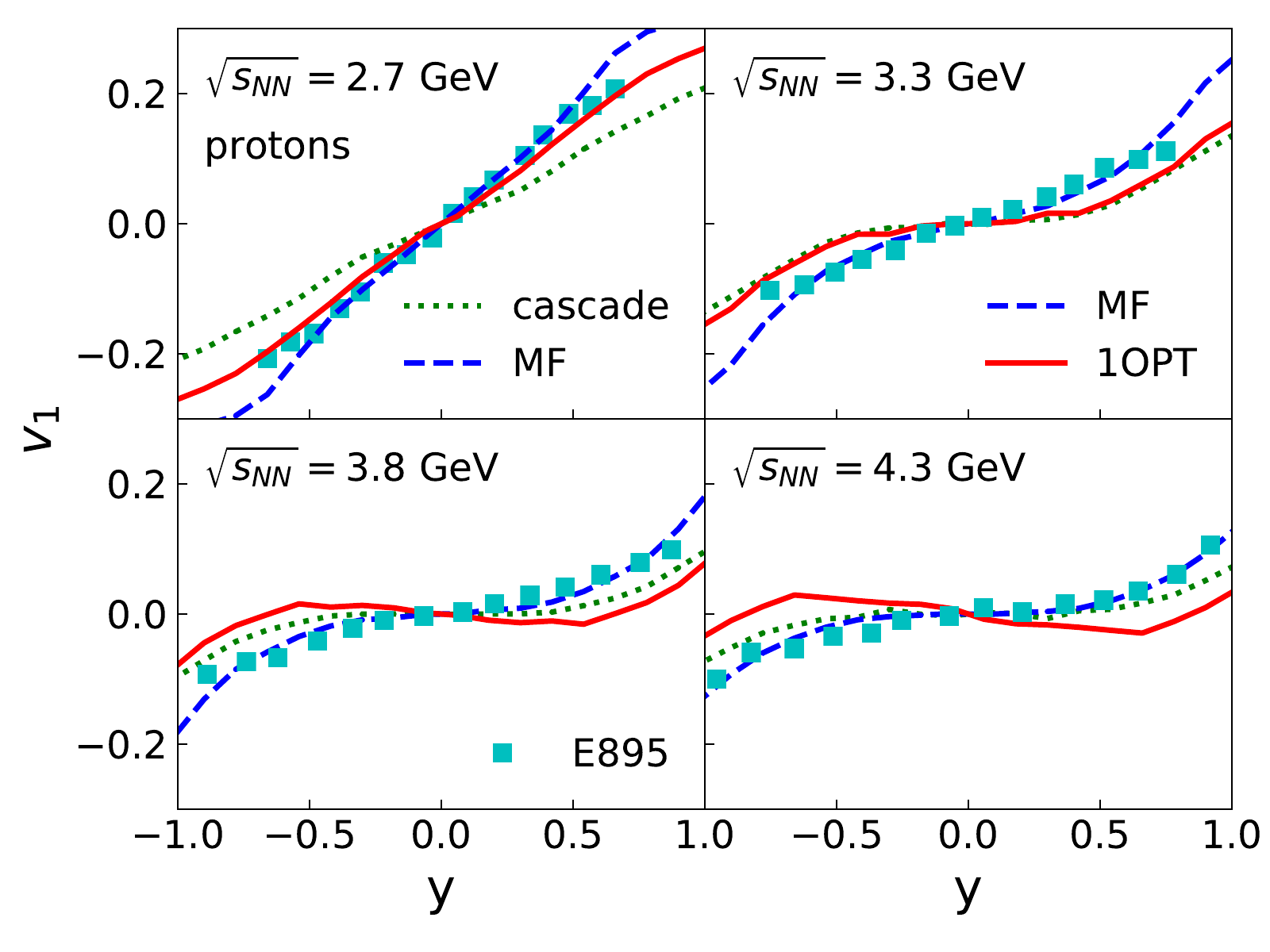}
   \caption{Directed flows  of protons as a function of rapidity
   in midcentral (10-40\%) Au+Au collisions
   at $\sqrt{s_{NN}}=2.7, 3.3, 3.8$ and 4.3 GeV
   from 
   JAM cascade (dotted lines), 
   JAM mean-field mode (JAM/MF) (dashed lines), 
   JAM with a first-order phase transition (JAM/1OPT) (solid lines)
   are compared with the data from the E895 Collaboration~\cite{E895v1}.
   }
   \label{fig:v1E985}
  \end{center}
\end{figure}

In Fig.~\ref{fig:v1E985}, rapidity dependence of the proton
directed flows from different modes of the model
are compared with the data~\cite{E895v1} at AGS energies.
The cascade mode always underestimates the slope of directed flow, which
indicates a lack of pressure generated by the cascade mode.
Inclusion of hadronic mean field generates more pressure and
improves the description of the data.
JAM/1OPT simulation predicts larger $v_1$ at $\sqrt{s_{NN}}=2.7$ GeV
than in the cascade mode,
which is because 1OPT EoS implemented in JAM
has a repulsive mean field in the hadronic phase.
While, as beam energy increases, the system hits the softest point
and the slope of proton-directed flow becomes negative
as originally predicted by hydrodynamical approach~\cite{DHRischke1995}.
However, experimental data show positive slope of protons,
and do not support a first-order phase transition
at AGS energies. As we will show in Fig.~\ref{fig:v2ex}
elliptic flows at AGS energies support the hadronic mean-field approach
as well, which predicts the suppressed elliptic flow.

\begin{figure}[th]
 \begin{center}
   \includegraphics[scale=0.5]{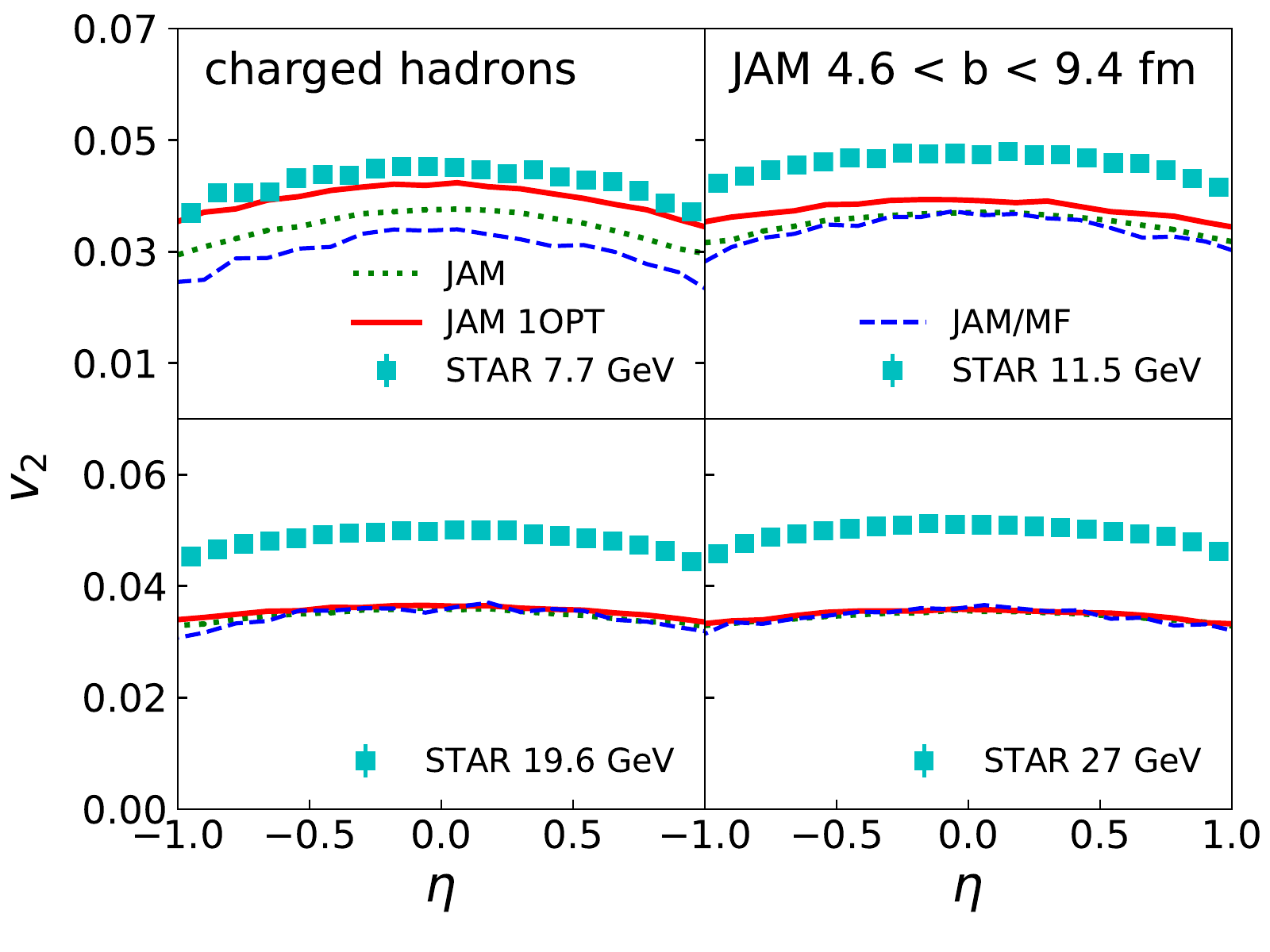}
   \caption{Elliptic flows of charged hadrons as a function of rapidity
   in midcentral (10-40\%) Au+Au collisions
   at $\sqrt{s_{NN}}=7.7, 11.5, 19.6$ and 27 GeV
   from 
   JAM cascade (dotted lines), 
   JAM with a fisrt-order phase transition (solid lines)
   are compared with the STAR data~\cite{STARv2}.
   }
   \label{fig:v2yFOch}
  \end{center}
\end{figure}

However, as seen in Fig.~\ref{fig:v2yFOch}, STAR data from
RHIC-BES experiments at $\sqrt{s_{NN}}=7.7$ GeV
seems to be consistent with a 1OPT scenario in JAM rather than
the result from JAM mean-field simulations:
A first-order phase transition enhances the elliptic flow,
while hadronic mean field suppresses the elliptic flow.
At higher beam energies, all JAM simulations predict the same
amount of elliptic flows which is below the data.
This may be due to the lack of
a partonic degree of freedom in the model.
It is reported that
the experimentally observed increase of the elliptic flow
with beam energy is reproduced by the inclusion of partonic interactions into
microscopic transport model 
PHSD~\cite{Konchakovski:2011qa,Konchakovski:2012yg}.

\begin{figure}[th]
 \begin{center}
   \includegraphics[scale=0.52]{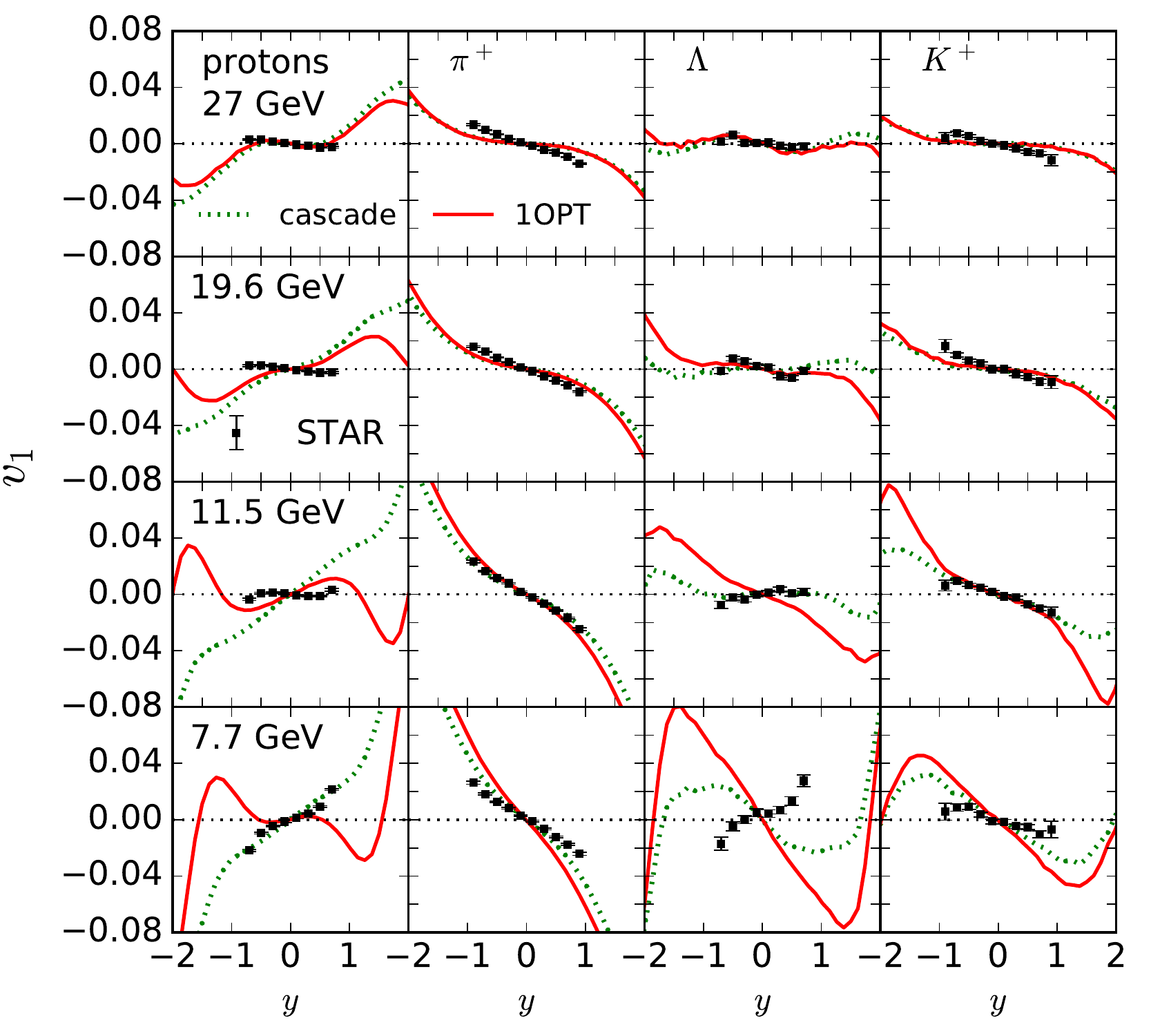}
   \caption{Rapidity dependence of directed flow
   for protons, pions, $\Lambda$s, and kaons
   in midcentral (10-40\%) Au+Au collisions
   at $\sqrt{s_{NN}}=7.7, 11.5, 19.6$ and 27 GeV
   calculated with
   the JAM cascade (dotted lines) and
   the JAM with a first-order phase transition (solid lines)
   are compared to STAR data~\cite{LAdamczykPRL2014,Adamczyk:2017nxg}.
   }
   \label{fig:v1star}
  \end{center}
\end{figure}

Recently, STAR measured directed flows of identified particles
including $\Lambda$s and kaons~\cite{Adamczyk:2017nxg}.
The measurements show that $\Lambda$ directed flow exhibits 
the same behaviour as protons, which cannot be accounted for
within our model as seen in Fig.~\ref{fig:v1star}.
JAM predicts positive slope for protons, and negative slope for $\Lambda$
below $\sqrt{s_{NN}}\leq 19.6$ GeV.
As we will investigate in detail a generation mechanism of
directed flow within our model, positive slope of proton-directed flow
is generated by meson-baryon scatterings: initial nucleon-nucleon
scatterings generate negative baryon directed flow.
Initially generated negative proton flow changes its direction
to the positive side by nucleon-pion scattering.
However, as the $\Lambda$-meson scattering rate is very small,
$\Lambda$ directed flow remains negative.
Thus, dynamical models  based on baryon and meson degrees of freedom
hardly describe the similarly of the directed flow between protons
and $\Lambda$s. We may need a model that incorporate
the effects of partonic interactions.
One expect to get the similar directed flow of protons and $\Lambda$s,
if their are mainly generated in the partonic phase.
Measurements of other baryons such as $\Xi$ and $\Omega$ baryons
may help to confirm the importance of deconfined phase
in the early stages of the reaction,
if they also show similar behavior as protons in directed flows.

The directed flows of pions and kaons exhibits negative slopes for all 
beam energies, which is consistent with our model results.
As we will study in detail the generation mechanisms of flows,
the main source of the negative slopes of meson flows are interaction
with spectator nucleons.


\section{hadronic re-scattering and spectator effects}
\label{sec:spec5}

Next we would like to investigate in detail the role of
spectator interactions on the flows.
As demonstrated in Ref.\cite{Nara:2017qcg},
the interplay between in- and out-of-plane flow
plays an essential role to determine the final strength of the elliptic flow. 
To perform a detailed analysis of the effects of 
hadronic rescattering and spectator matter on anisotropic flows,
we compute anisotropic flows
by disabling meson-baryon ($MB$) and meson-meson ($MM$) scatterings;
a simulation in which 
the cross sections of $MB$ and $MM$ scatterings are set to be zero.
The effect of spectator shadowing is studied by disabling the
interaction between spectator nucleons and participants,
where spectator nucleons are defined in our approach
as the nucleons that are not in the collision list
of the initial state nucleon-nucleon collisions.
More specifically, we first compute possible nucleon-nucleon collisions
after sampling nucleons inside two nuclei and boost them,
which corresponds to the initial Glauber-type nucleon-nucleon collisions.
If nucleons in the projectile nucleus are considered not to collide with
any other nucleons in the target nucleus,
these nucleons are regarded as spectator nucleons, which are initially located
outside a reaction zone of two nuclei.
We assume that collision can happen
when the impact parameter of the two incoming particles $b$
are inside the interaction distance specified by the
geometrical interpolation of the cross section $\sigma$, i.e.,
$b \leq \sqrt{\sigma/\pi}$.
Of course, if beam energy is not very high, ``spectator nucleons"
can interact with nucleons in participant zone.
In our simulations, we update collision list after every two-body collision,
thus initially predicted nucleon-nucleon collision can be modified
according to the dynamics of the collision of two nuclei.

\subsection{Directed flow}

\begin{figure}[t]
 \begin{center}
   \includegraphics[scale=0.45]{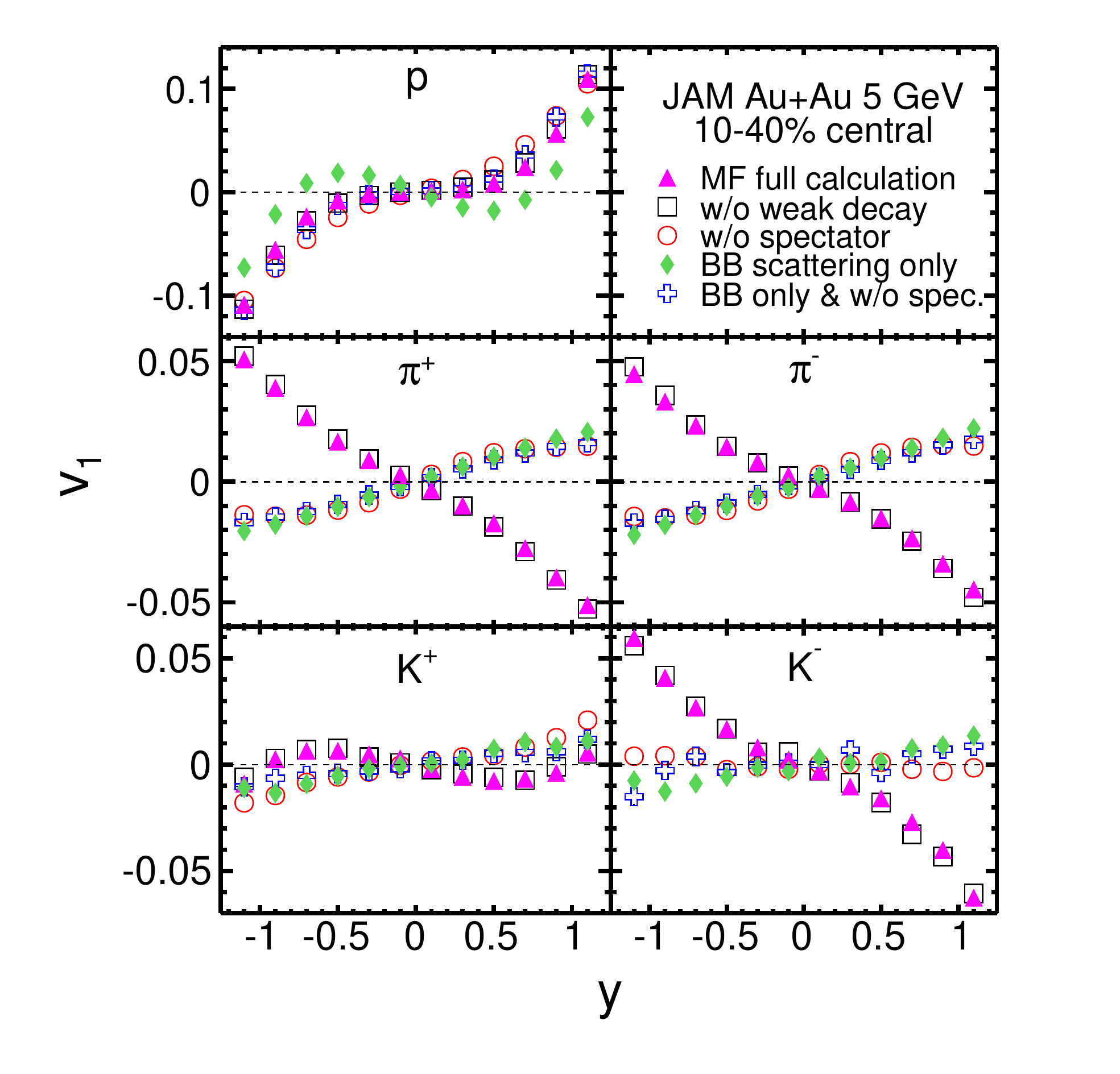}
   \caption{Directed flows as a function of rapidity
   in midcentral (10-40\%) Au+Au collision at $\sqrt{s_{NN}}$=5 GeV 
   from JAM mean-field mode (JAM/MF) (triangles), 
   JAM/MF without weak decay (squires),
   JAM/MF without spectator  (circles),
   JAM/MF with only baryon-baryon collisions  (diamonds),
and JAM/MF with only baryon-baryon collisions and without spectator
  (crosses).
   The left and right panels show the results for identified particles
   $p$, $\pi^{+}$, $K^{+}$ and antiparticles
   $\pi^{-}$, $K^{-}$, respectively.}
   \label{fig:v1m5}
  \end{center}
\end{figure}

Fig.~\ref{fig:v1m5} shows the rapidity dependence of directed flow
for identified particles ($p$, $\pi^{+}$, $K^{+}$) and 
corresponding antiparticles ($\pi^{-}$, $K^{-}$) 
in midcentral (10-40\%) Au+Au collision at $\sqrt{s_{NN}}=5$ GeV 
from JAM mean-field (JAM/MF) simulations.
JAM/MF simulation predicts positive $v_1$ for protons and negative
$v_1$ for pions and kaons.
To see the effect of $MB$ and $MM$ scattering,
we show the results of simulation by switching off
$MB$ and $MM$ scatterings
i.e., baryon-baryon ($BB$) collision only (diamonds) simulations,
which yield negative $v_1$ for protons
and positive $v_1$ for pions and kaons.
We also test the effects of spectator matter by disabling
the interaction with spectator [w/o spectator (circles)].
This simulation shows that nucleon $v_1$ is positive
and larger at midrapidity compared to the full simulations.
The Pion and kaon $v_1$ are also positive with this simulations.
If we disable both interaction with spectator and meson-baryon scatterings,
nucleon $v_1$ becomes almost zero at midrapidity.
Thus, it is clear that negative nucleon $v_1$ in the $BB$ scattering
only simulation is due to the shadowing effect by the spectator matter,
whereas meson-baryon collisions reflect nucleons to the positive $v_1$
and pions to the negative side.
It is also demonstrated that \textit{all} of the negative $v_1$ for pions
are generated by the interaction
between spectator nucleons and pions.

We have also studied the effects of weak decays since
some nucleons and pions are produced
from the weak decay such as $\Lambda^0$, and $\Sigma^-$,
and this may affect the distribution of flows.
We observe that directed flow is not sensitive to the weak decay effects
at 5 GeV.

\subsection{Elliptic flow}

\begin{figure}[t]
 \begin{center}
   \includegraphics[scale=0.45]{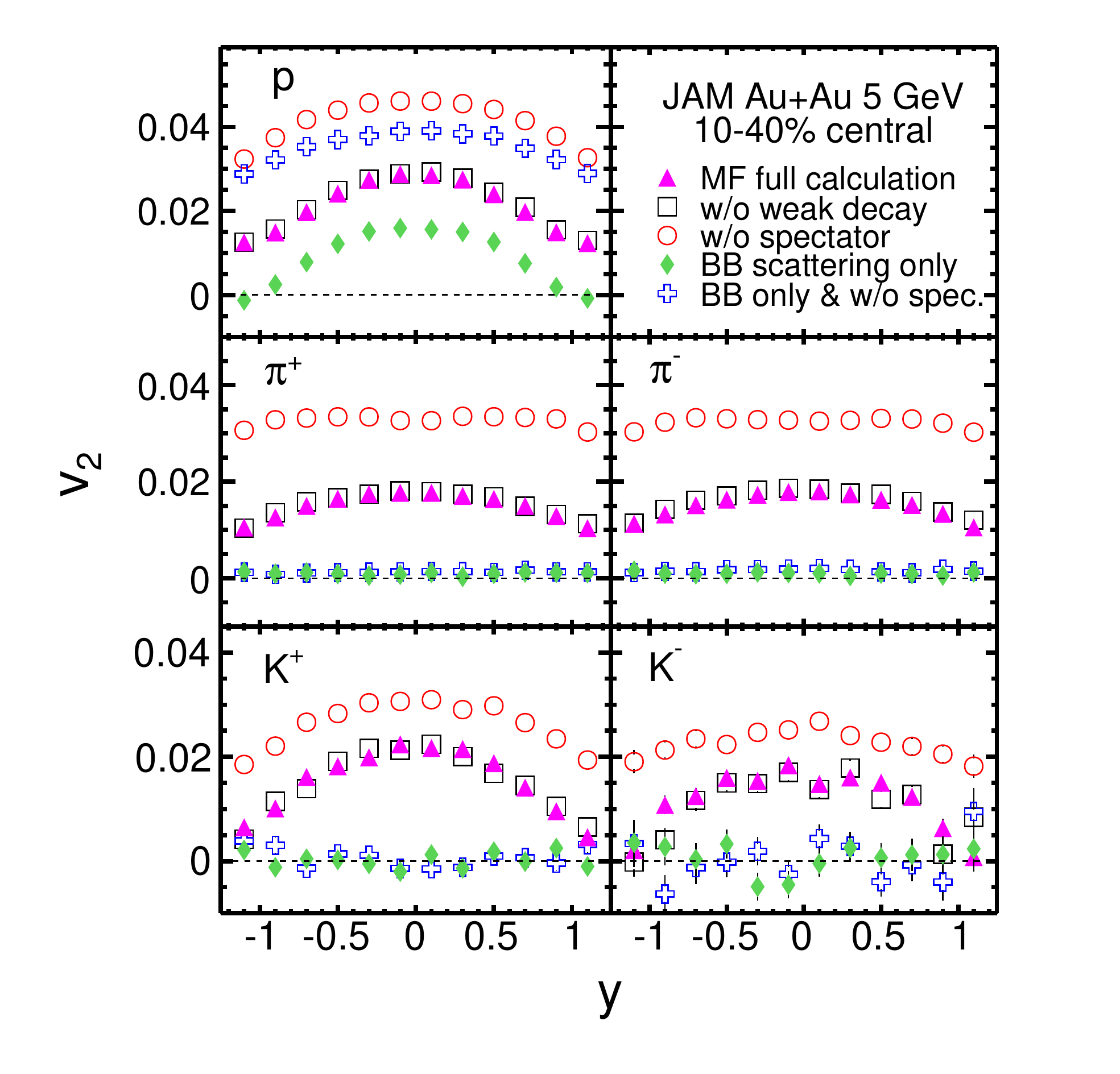}
   \caption{Same as Figure~\ref{fig:v1m5}, but for $v_{2}$.}
 \label{fig:v2m5}
  \end{center}
 \end{figure}

We do the same excises for elliptic flow.
In Fig.~\ref{fig:v2m5}, the rapidity dependence of $v_{2}$
for identified particles ($p$, $\pi^{+}$, $K^{+}$)
and corresponding antiparticles ($\pi^{-}$, $K^{-}$)
in midcentral Au+Au collisions at $\sqrt{s_{NN}}=5$ GeV
from JAM mean-field mode are presented.
Elliptic flow is generated by the scattering among hadrons in JAM.
When $MB$ and $MM$ scatterings are disabled,
nucleon $v_2$ are smaller by about 20\%,
and pion and kaon elliptic flows are zero.
It is seen that the effect of spectator on the elliptic flow 
is very large for entire rapidity range at 5 GeV for all particles.
The spectator effects on the elliptic flow for nucleons and pions
are 20-40 \% reduction at $\sqrt{s_{NN}}=5$ GeV at midrapidity.
The shape of rapidity distribution is mainly determined by
the degree of suppression of the elliptic flow.

\begin{figure}[!htb]
 \begin{center}
   \includegraphics[scale=0.45]{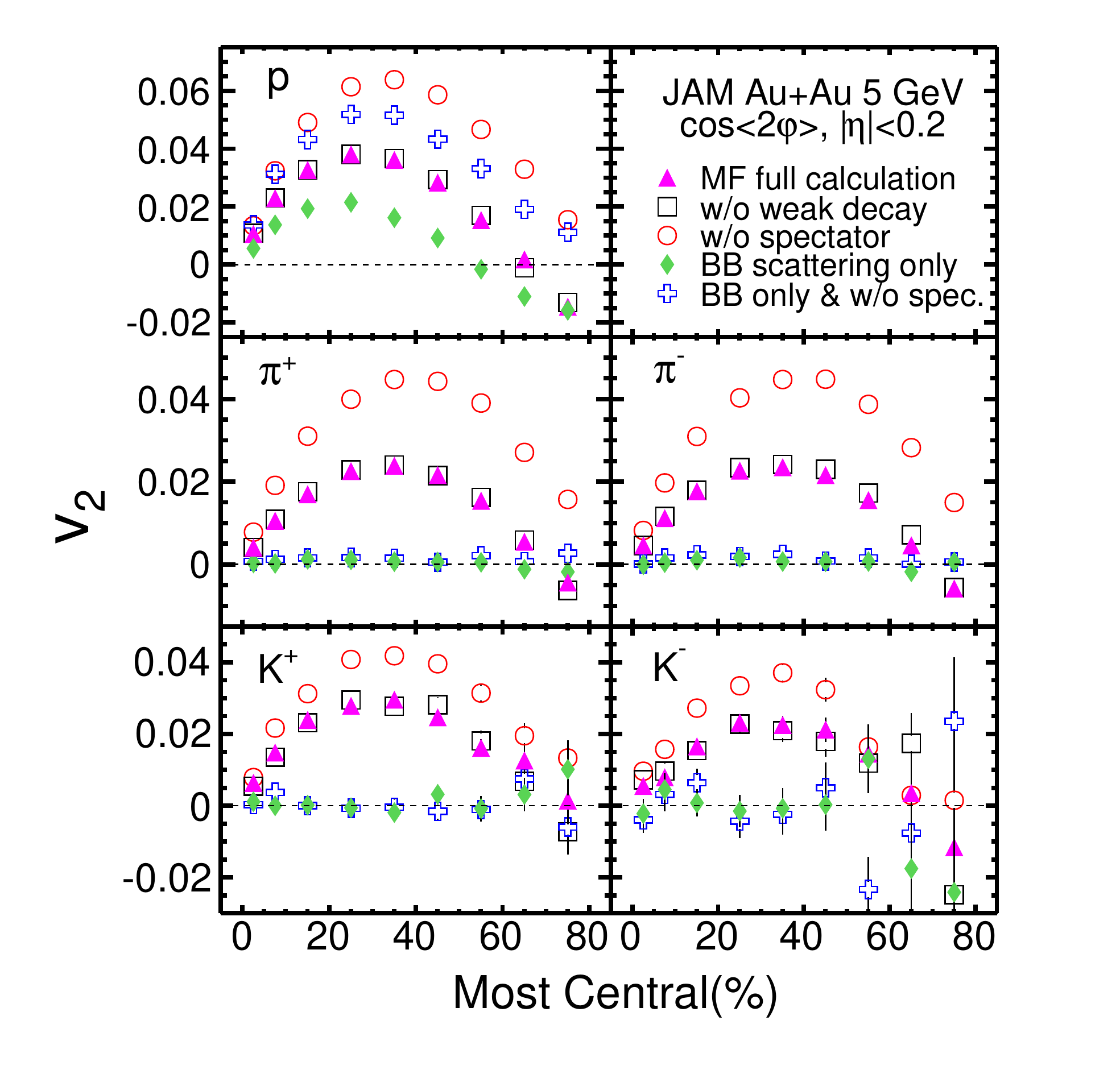}
   \caption{Same as Figure~\ref{fig:v2m5},
   but for the centrality dependence of
   the $\eta$ integrated $v_{2}$
   at $|\eta| < 0.2$.
   }
   \label{fig:v2cent}
   \includegraphics[scale=0.4]{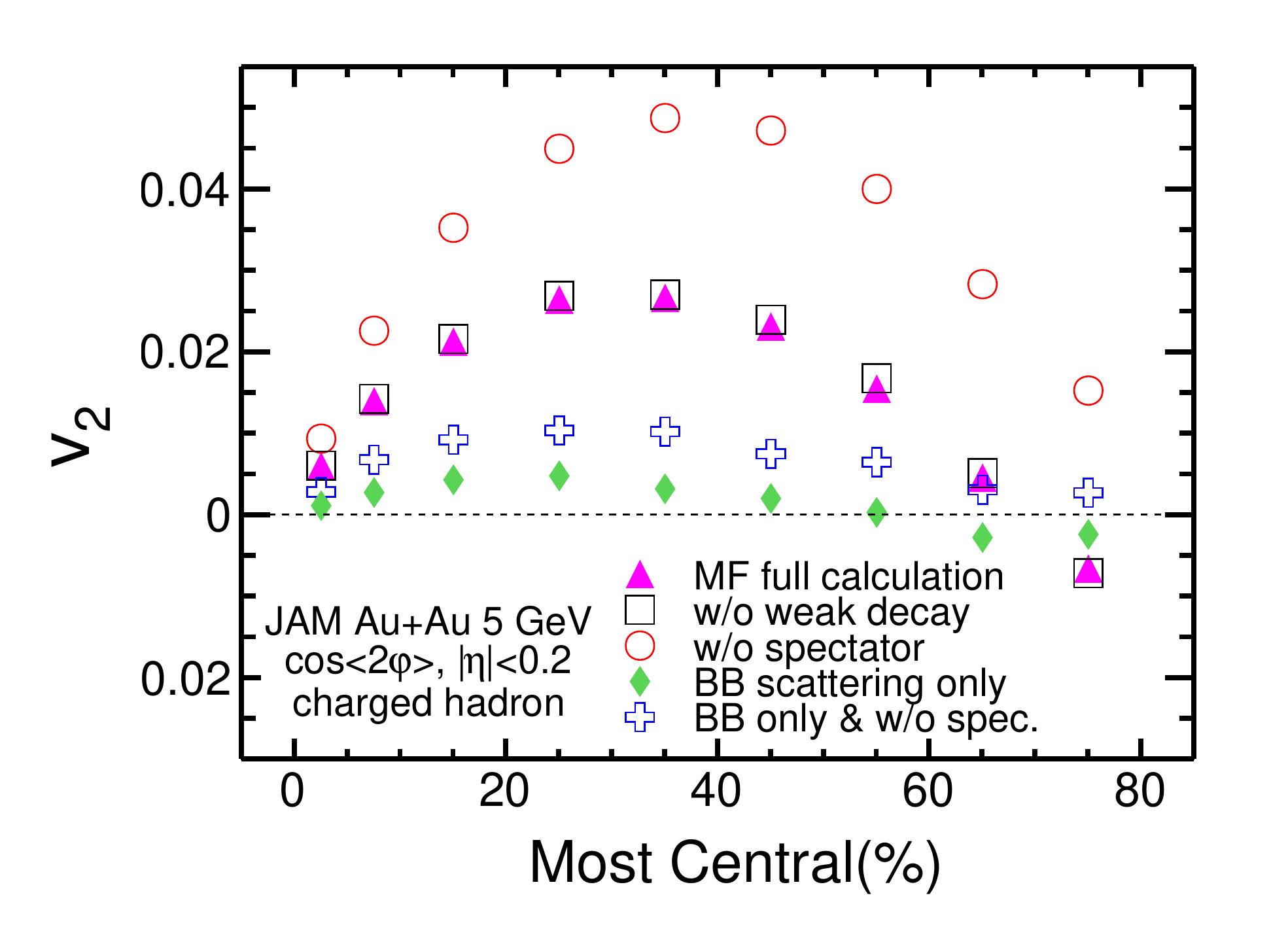}
   \caption{Same as Fig.~\ref{fig:v2cent} but for charged particles.}
   \label{fig:v2centch}
  \end{center}
 \end{figure}
 
Collision centrality dependence may also contain important information
about the collision dynamics in heavy-ion collisions.
Figs.~\ref{fig:v2cent} and \ref{fig:v2centch} display
the centrality dependence of integrated $v_{2}$ for particles
($p$, $\pi^{+}$, $K^{+}$, $\pi^{-}$, $K^{-}$)
and charged hadrons from JAM mean-field simulations
in Au+Au collisions at $\sqrt{s_{NN}}=5$ GeV.
The suppress of elliptic flow by the spectator can be seen
for all centrality for all particles except for very central collisions.
It is interesting to see that elliptic flow becomes
negative at very peripheral collisions even at 5 GeV.

\subsection{squeeze-out and softening}

In this section, we discuss the squeeze-out effect on the elliptic flow
when a softening of EoS happens.
In Refs.~\cite{Chen:2017cjw,Nara:2017qcg},
it was found that softening of EoS leads to an enhancement of elliptic flow,
which is considered to be caused
by the suppression of squeeze-out at high baryon density region,
and proposed as a possible signature of a first-order phase transition.
Attractive orbit simulation shows the enhancement of $v_2$
for pions~\cite{Chen:2017cjw}, 
whereas a simulation with a first-order phase transition
predicts enhancement of $v_2$ for both protons and pions~\cite{Nara:2017qcg}.
Here we shall see explicitly
that the softening indeed suppresses the squeeze-out.

\begin{figure}[!htb]
 \begin{center}
   \includegraphics[scale=0.75]{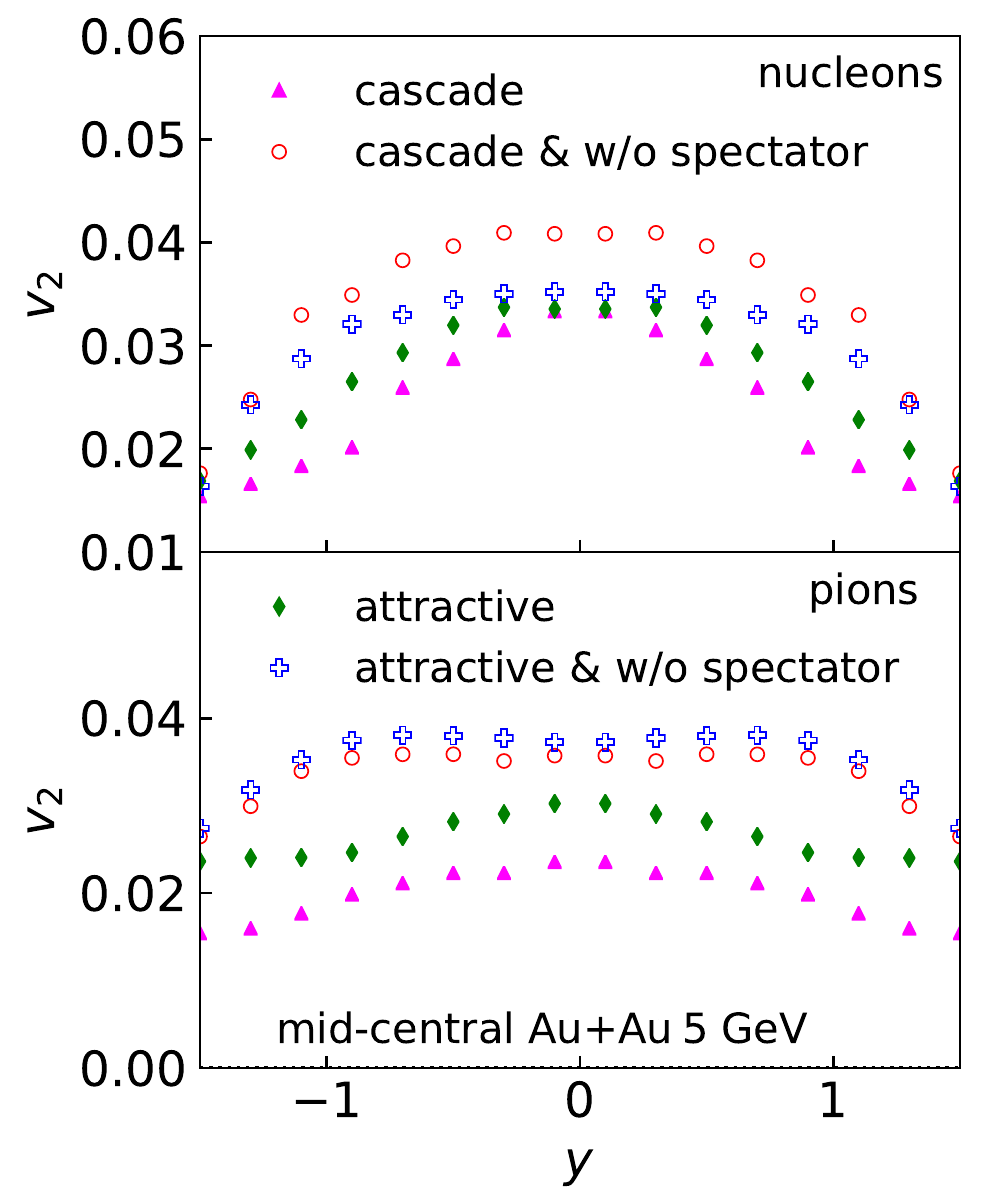}
   \caption{Rapidity dependence of $v_2$ for nucleons (upper)
   and pions (lower) in midcentral Au+Au at $\sqrt{s_{NN}}=5$ GeV
   are compared with the JAM cascade and JAM attractive orbit mode
   with and without spectator interaction.}
   \label{fig:v2att5}
  \end{center}
\end{figure}

To understand the role of squeeze-out in the case of softening
more qualitatively,
the elliptic flow from attractive orbit simulation
is compared with the standard JAM cascade simulation
with and without spectator in Fig.~\ref{fig:v2att5}.
Without spectator interaction,
attractive orbit simulation yields less nucleon $v_2$
than from standard cascade simulation.
This is because attractive orbit simulation
leads to very small pressure for all space-time regions of the reaction.
However, situation changes in the case of the presence of a spectator.
It blocks the in-plane expansion at early times and suppress $v_2$. 
The degree of suppression is weaker in the case of softening.
We observe that the effect of squeeze-out becomes less in the case of
softening: a reduction of nucleon $v_2$ in the cascade mode is about 30\%,
while there is almost no reduction in the attractive orbit mode.
In the case of pions, $v_2$ in the attractive orbit simulation
is almost the same as that of cascade simulation
without interaction with the spectator matter.
It is seen that pion $v_2$ are suppressed more
by the spectator for the cascade simulation.
The net effect is a larger $v_2$ in the attractive mode compared to
the cascade mode.

\begin{figure}[!htb]
 \begin{center}
   \includegraphics[scale=0.75]{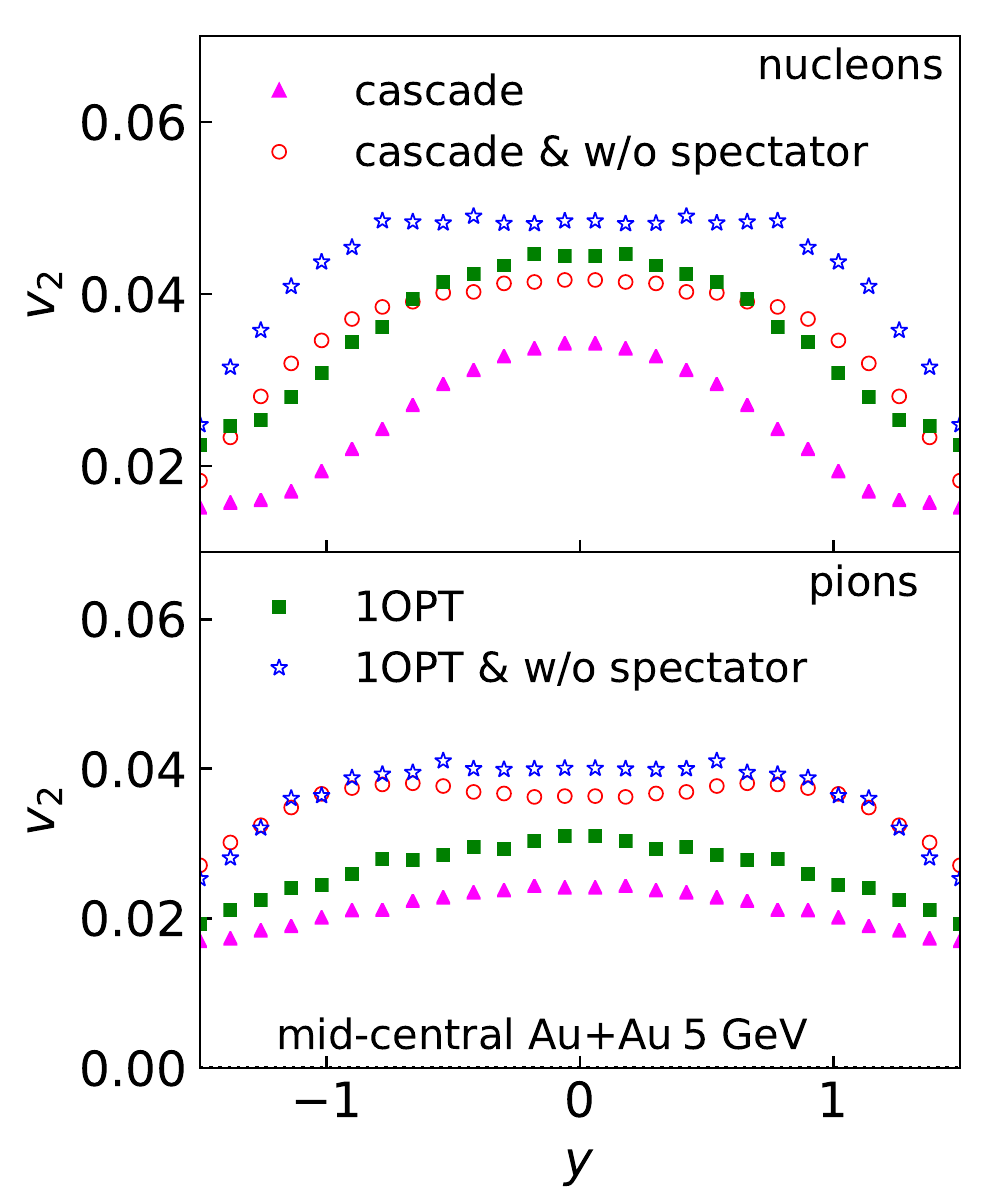}
   \caption{Rapidity dependence of $v_2$ for nucleons (upper)
   and pions (lower) in midcentral Au+Au at $\sqrt{s_{NN}}=5$ GeV
   are compared with the JAM cascade and JAM first-order phase transition
    mode with and without spectator interaction.}
   \label{fig:v2FO5}
  \end{center}
\end{figure}

Attractive orbit simulation strongly suppresses the pressure of the system
for whole reaction time and all spatial region of the system
regardless of its energy density
since we force the attractive orbit for
all two-body scatterings without any restrictions, which
is considered to be a maximum effect of softening within our approach.
To take into account the softening effect only if
the system enters the softening point,
we perform a simulation with a 1OPT~\cite{Nara:2016hbg},
and the results are shown in Fig.~\ref{fig:v2FO5}.
The elliptic flow in the 1OPT simulation
without spectator becomes larger than the cascade results.
This may be because that the system is compressed more
in the 1OPT simulation, and as shown in Ref.~\cite{Nara:2017qcg},
initial eccentricity becomes larger, but during the
expansion, the system eventually goes out from the soft region,
and generates stronger in-plane flow.
We also note that nucleon elliptic flow
is not suppressed by the spectator shadowing at midrapidity
in the case of 1OPT as well, while
pion elliptic flow is suppressed by the spectator even 
in the 1OPT for pions, although the degree of suppression
is less.

These analysis indicates that final value of elliptic flow
at high baryon density region
is determined by the interplay between in-plane and out-of-plane
emission, and it is very sensitive to the pressure of the system.

\section{Beam energy dependence}
\label{sec:beam}

The beam energy dependence of the spectator effects will be investigated
in this section.
At sufficiently high energies such as at top RHIC and LHC energies,
hybridization of reaction dynamics is successful in describing the
collision of high energy heavy-ion collisions.
One of them is the ``factorization'' of reaction time in which 
heavy-ion collision can be separated by the initial Glauber type
nucleon-nucleon collisions that provides the initial conditions
for the subsequent space-time evolution of the system by, e.g. hydrodynamics
or transport theories.
One may estimate the minimum beam energy at which
this factorization of reaction time becomes applicable
by considering a passing time of two nuclei.

The passing time of two nuclei $t_\mathrm{pass}$ can be estimated
by using the radius $R$, velocity $v$ of the nucleus,
and $\gamma$ factor as
$t_\text{pass}=2R/\gamma/v \approx 0.9$ fm/$c$ at 27 GeV.
Thus we expect that collision dynamics changes around this beam energy,
assuming that typical formation time of produced particle is about 1 fm/c.
Above this energy, the initial condition for a subsequent
evolution of the system can be obtained by the initial particle production
just after the collision of two nuclei, which may be computed by
Glauber model or theories based on the color glass condensate (CGC)
~\cite{Heinz:2013th, Gale:2013da,Hirano:2004en,Hirano:2012kj}.
On the other hand, below this beam energies, this factorization 
breaks down and we need to take into account the rescatterings
of particles during the passage of two nuclei.
A new dynamical initialization is proposed in the hydrodynamical
approach to simulate lower beam energies in Ref~\cite{Shen:2017bsr}
talking into account such effects.
If rescatterings among produced particles happen during the passage
of two nuclei, interactions with the spectator nucleons become
also important in the early stages of heavy-ion collision, and
it has a significant impact on the anisotropic flows.
 
When we go down to even lower beam energies
such as $\sqrt{s_{NN}}\lesssim 5$ GeV,
nuclear mean-field effects become important,
and microscopic transport approaches based on
the Boltzmann-Uehling-Uhlenbeack (BUU) equation~\cite{BUU} or
the quantum molecular dynamics (QMD) $N$-body phase-space dynamics~\cite{QMD}
have been successful in describing collision dynamics.
We shall study the transition of generation mechanisms 
of anisotropic flows within
a microscopic transport approach below.

\subsection{AGS energies}

\begin{figure}[!htb]
 \begin{center}
   \includegraphics[scale=0.60]{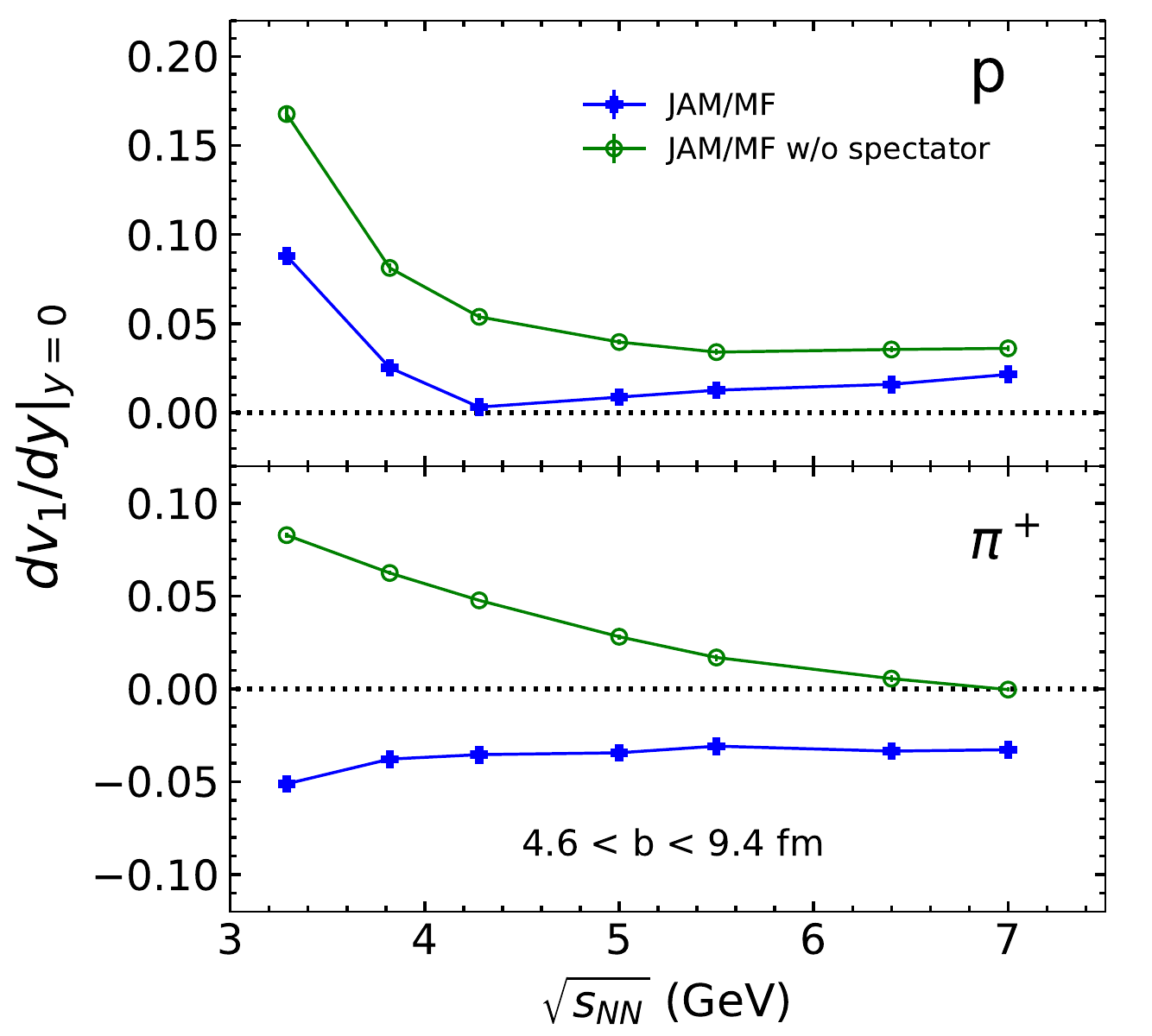}
   \caption{Beam energy dependence of the slope of directed flow
   in midcentral (10-40\%) Au+Au collision
   from JAM mean-field mode (MF) (crosses),
and MF without spectator  (circles).
   Slope $dv_1/dy$ is obtained by the cubic fit at the range of
   $|y|<0.8$.
   The top and bottom panels show the results for 
   protons and positive pions,
   respectively.}
   \label{fig:ene_v1}
  \end{center}
\end{figure}

We first investigate beam energy dependence at AGS energies where
nuclear mean-field effects play important role.
In Fig.~\ref{fig:ene_v1},the beam energy dependence of
the slopes of directed flow $dv_1/dy$ at midrapidity are shown
for protons and positive pions
from JAM with the mean-field simulations.
The $v_1$ slope at midrapidity without spectator interaction
is larger compared to the full calculations.
Thus, it demonstrates that
directed flow at midrapidity is significantly influenced by
the shadowing effect by the spectator even at midrapidity.
At higher beam energies, as we shall investigate in the next
section, the shadowing effect by the spectator is not seen
for the nucleon $v_1$ slope
because of increasing number of scatterings between pions and nucleons.
The $v_1$ slopes for pions are negative in the full simulations,
while they are positive without spectator interactions.
Thus, negative-directed flow for pions is generated solely by
the interaction with the spectator in this beam energy range.

\begin{figure}[!htb]
 \begin{center}
   \includegraphics[scale=0.60]{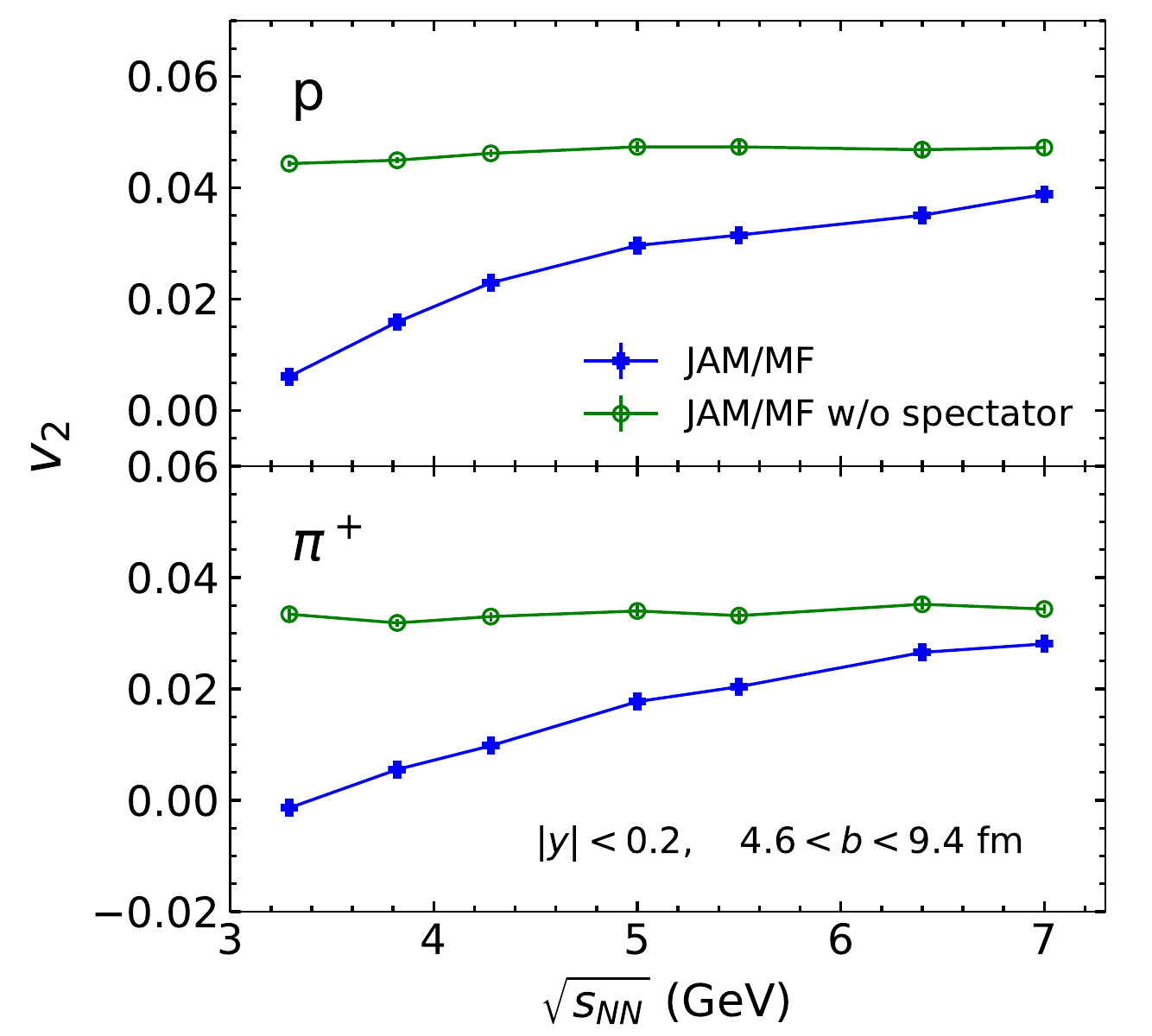}
   \caption{Same as Fig.~\ref{fig:ene_v1},
   but for beam energy dependence of elliptic flows $v_{2}$
   at midrapidity $|y|<0.2$.
   }
   \label{fig:ene_v2}
  \end{center}
\end{figure}

In Fig.~\ref{fig:ene_v2},
beam energy dependence of the elliptic flows of protons and positive pions
at midrapidity $v_2(|y|<0.2)$ is shown.
Without spectator interaction, elliptic flow excitation function
is almost flat for both protons and pions
at beam energies $3<\sqrt{s_{NN}}<7$ GeV
which indicates that elliptic flow at midrapidity in this energy region
is mainly determined by the degree of shadowing by the spectator.
The relative contribution of the squeeze-out effect to the elliptic flow 
decreases smoothly as the beam energy increases:
the out-of-plane flow (squeeze-out) is large
in the beam energy $\sqrt{s_{NN}}<4$ GeV,
while both out-of plane and in-plane contributions
are important at $5 < \sqrt{s_{NN}}<6$, and then,
in-plane flow becomes dominant at $\sqrt{s_{NN}}> 6.5$ GeV.

\subsection{RHIC-BES energies}

We now study the beam energy dependence of 
the effects of rescattering and spectator shadowing
for directed and elliptic flows at RHIC-BES energy region.

\subsubsection{Directed flow}

\begin{figure}[!htb]
 \begin{center}
   \includegraphics[scale=0.78]{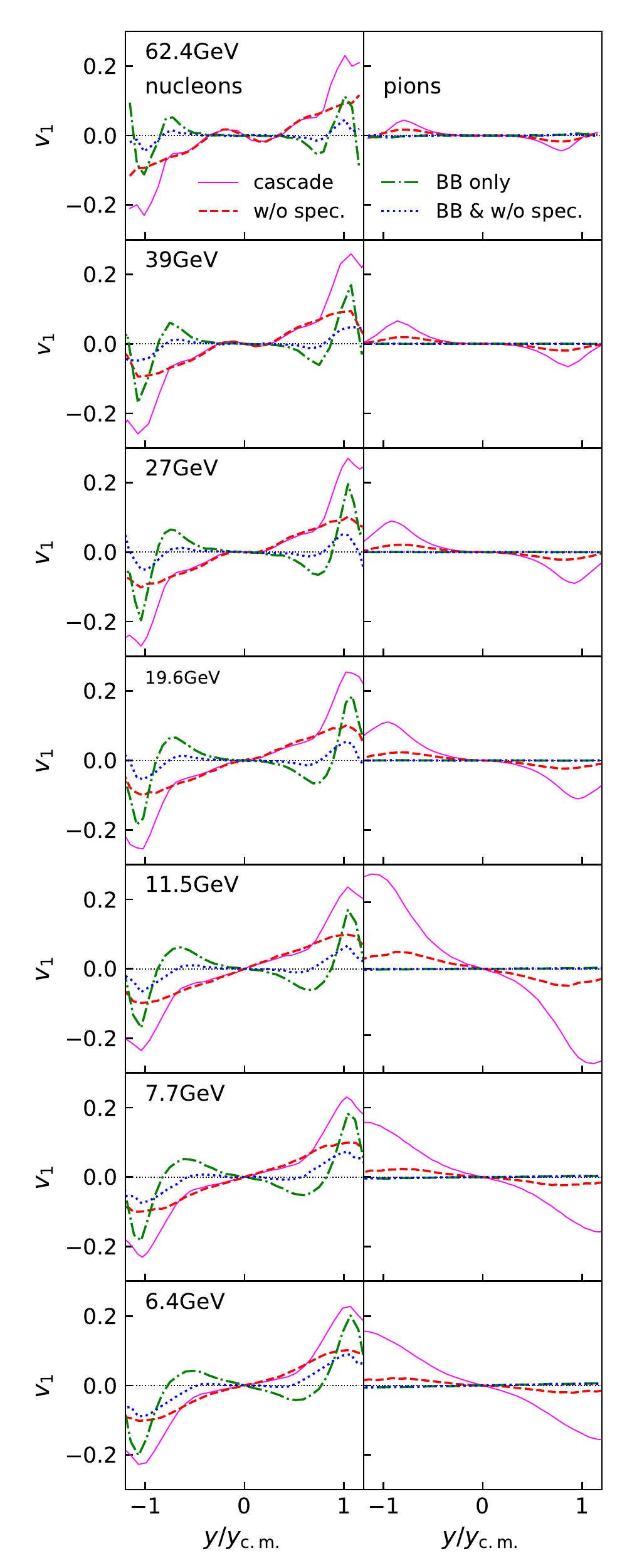}
   \caption{Rapidity dependence of directed flow for nucleons (left panel)
   and pions (right panel) are compared in midcentral Au + Au
collisions ($4.6 < b < 9.6$fm) at $\sqrt{s_{NN}}$  = 6.4, 7.7, 11.5, 19.6, 27, 39 and 62.4 GeV       from JAM cascade simulations. In the calculations,
   acceptance cut $0.4\leq p_T\leq 2$ GeV/$c$ for nucleons,
   $0.2\leq p_T\leq 1.6$ GeV/$c$ for pions
   are imposed.
 }
   \label{fig:v1ex}
 \end{center}
\end{figure}

Fig.~\ref{fig:v1ex} shows the beam energy dependence of
directed flow as a function of rapidity
for nucleons (left panel) and pions (right panel)
from 6.4 to 62.4 GeV obtained by the cascade mode.
One may refer Refs.~\cite{YNaraPRC2016,YNaraNPA2016}
for the comparison with the STAR data at midrapidity with different EoS.
Here we focus on the rapidity dependence on the flows up to
the beam rapidity region.

Let us first compare the simulation without rescattering; i.e.,
simulations which include only $BB$ collisions.
The shadowing effect by the spectator can be seen in the
$BB$ only simulations up to 19.6 GeV.
Namely, the slope of nucleon-directed flow is negative
when $MB$ and $MM$ scatterings are disabled (dotted-dashed lines),
while it is almost zero
if there are only $BB$ scatterings and there is no spectator (dotted lines).
This implies that nucleon-directed flow is negative due to
the shadowing effect by the spectator when only $BB$ collisions are
included. Once $MB$ and $MM$ scatterings
are turned on, slope  becomes positive, and
the shadowing effect is washed out:
the directed flow of nucleon $v_1$ at $y/y_\mathrm{c.m.}<0.75$
in the full cascade simulation with spectator (solid lines)
is almost identical to the results of
the simulations without spectator interaction (dashed lines).
Thus the shadowing effect on the directed flow of nucleons is
not seen in the final strength of the directed flow
at the rapidity region of $y/y_\mathrm{c.m.}<0.75$.
In contrast, most of the pion-directed flow is generated by the interaction
with the spectator matter.

When beam energy becomes higher than 20 GeV,the collision dynamics
changes dramatically, as expected from the passing time argument.
The slope of the nucleon-directed flow at midrapidity
becomes negative from 27 GeV in the full cascade simulations,
as predicted by Ref.~\cite{RJMSnellingsPRL2000} called wiggle shape.
It is explained~\cite{RJMSnellingsPRL2000}
by the space-momentum correlation and the degree of baryon stopping,
which is not a shadowing effect by the spectator matter.
This is confirmed by the results that
$BB$ only simulation yields the same results when interaction
with the spectator matter is turned off: the nucleon-directed
flow is almost zero for both $BB$ only simulation and $BB$ only without 
spectator simulation at high energies.
In our hadronic transport approach,the space-momentum correlation is generated
by rescattering among particles after two nuclei pass through each other
at high energies above 30 GeV,
thus $MB$ scattering
generate a negative nucleon-directed flow at high energies.
It is shown that
tilted initial condition yields the negative $v_1$ slope
in hydrodynamics at RHIC~\cite{Bozek:2010bi}.
We note that
CGC also predicts tilted initial conditions~\cite{Adil:2005bb}.

It is interesting to observe that
at forward rapidity region  $y/y_\mathrm{c.m.} >0.75$,
the directed flow of nucleon looks identical
for all beam energies suggesting the universal mechanism to
generate directed flow at forward rapidity regions.
At high energies,the pion-directed flow at forward rapidity region 
is also generated by the spectator shadowing.

\begin{figure}[!htb]
 \begin{center}
   \includegraphics[scale=0.45]{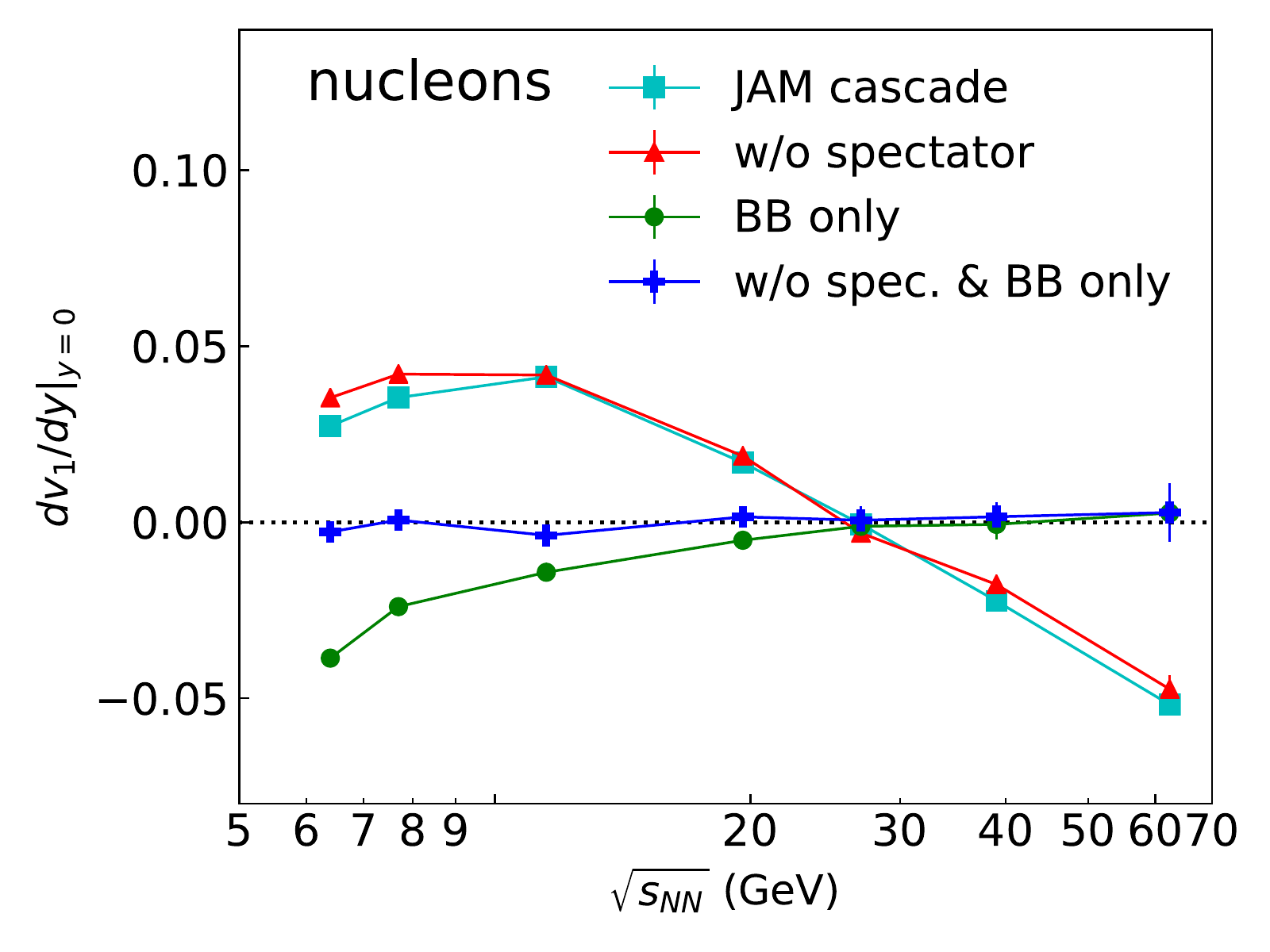}
   \caption{Beam dependence of the slope of nucleon directed flow
   in midcentral (10-40\%) Au+Au collision
   from JAM cascade (squires),
   cascade without spectator (triangles), 
   cascade with only baryon-baryon collisions  (circles),
and cascade with only baryon-baryon collisions and without spectator
  (crosses).}
   \label{fig:v1slopeN}
  \end{center}
\end{figure}

To see more clearly the different mechanisms for the origin of
directed flow of nucleons at midrapidity, slopes of nucleon
$v_1$ is depicted in Fig.~\ref{fig:v1slopeN} as a function of
beam energy, where the slope is obtained by the linear fit at $|y|<0.5$.
If both spectator interaction and $MB$ and $MM$ scatterings are disabled,
the directed flow is not generated (crosses).
Once interaction with the spectator is included, $BB$  collisions
generate the negative-directed flow as a result of the shadowing up to
the beam energy of around 30 GeV,
while above 30 GeV, the $BB$ collision does not generate directed flow
even with spectator. Thus there is no effect of spectator shadowing 
on the directed flow of nucleon at midrapidity above 30 GeV.
The role of final state interactions (mostly $MB$ and $MM$ scattering
in our case) to the slope of nucleon-directed flow is opposite
at low and high energies.
Namely, the effect of $MB$ and $MM$ collisions
is  to generate positive nucleon directed flow below 30 GeV, while
they generate negative-directed flow above 30 GeV.
Thus the dynamical origin of directed flow changes at 30 GeV.

\subsubsection{Elliptic flow}

\begin{figure}[!htb]
 \begin{center}
   \includegraphics[scale=0.76]{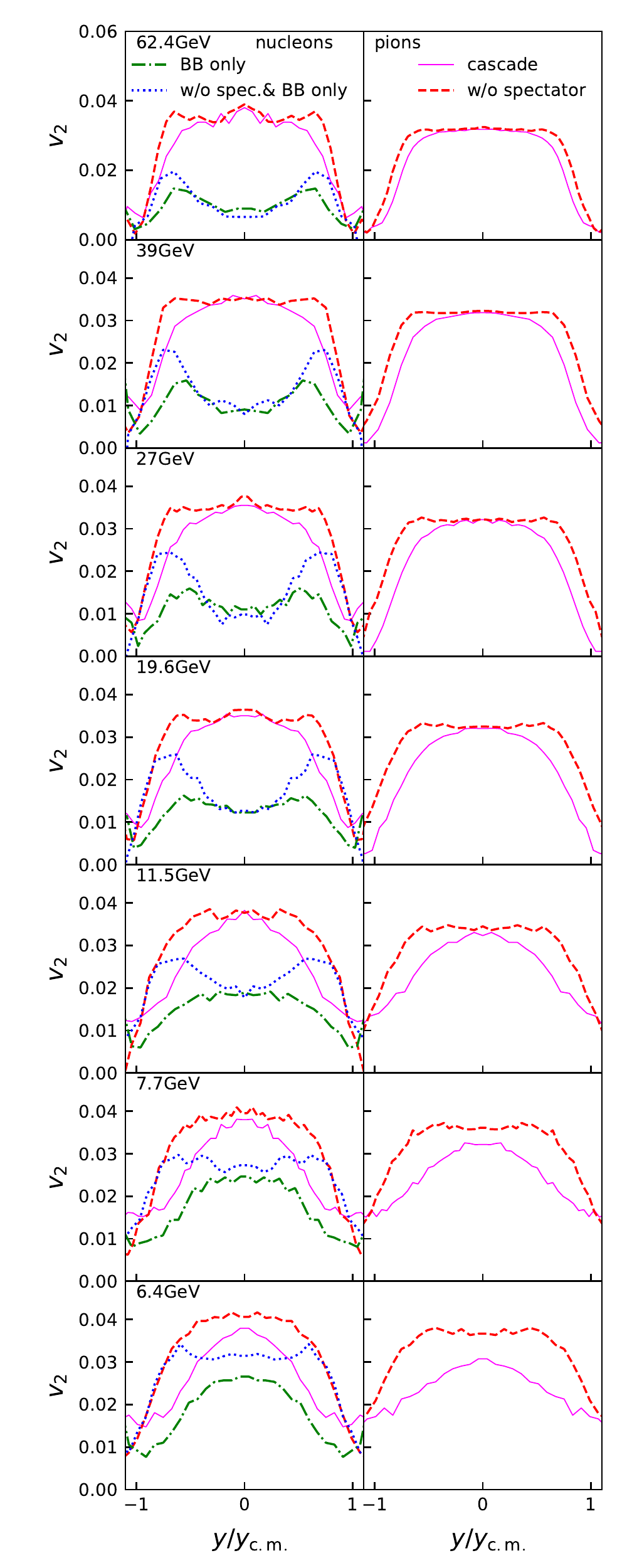}
   \caption{Rapidity dependence of $v_2$ for nucleons (left panel)
   and pions (right panel)  are compared in midcentral Au + Au
collisions ($4.6 < b < 9.6$fm) at $\sqrt{s_{NN}}$  = 6.4, 7.7, 11.5, 19.6, 27, 39 and 62.4 GeV from JAM cascade simulations. In the calculations,
   acceptance cut $0.4\leq p_T\leq 2$ GeV/$c$ for nucleons,
   $0.2\leq p_T\leq 1.6$ GeV/$c$ for pions
   are imposed.
 }
   \label{fig:v2ex}
 \end{center}
\end{figure}

In Fig.~\ref{fig:v2ex}, rapidity dependence of the elliptic flow
for nucleons and pions are plotted for the beam energy range of
$6.4\leq\sqrt{s_{NN}}\leq62.4$ GeV in midcentral Au+Au collisions.
It is seen by comparing the result of JAM cascade (solid line)
with the simulation without spectator interaction (dashed line) that
at lower energies $\sqrt{s_{NN}} \leq 7.7$ GeV,the suppression of
elliptic flow by the spectator matter is large for all rapidity region for
both nucleons and pions.
The contribution of $MB$ and $MM$ scatterings to the elliptic flow
is increasingly significant as the beam energy increases.
The spectator effect on the elliptic flow at midrapidity 
disappears at above 11.5 GeV, but the suppression of elliptic flow
is still seen at $y/y_\text{c.m.}>0.5$, which becomes smaller
as beam energy increases.
At higher beam energies,the elliptic flow for both nucleons and pions
becomes flat in hadronic cascade simulation.
It was reported that at forward rapidity region,
the elliptic flow from hadronic transport models is
consistent with the RHIC data~\cite{Bratkovskaya:2003ux}.
However, three-dimensional-hydrodynamics supplemented by a hadron transport after burner 
also predicts a compatible amount of elliptic flow at forward rapidities
\cite{Hirano:2005xf,Nonaka:2006yn}.
Our results demonstrate that
the interaction with spectator plays a minor role
for the generation of elliptic flow
for a wide range of rapidity region at sufficiently higher energies.

\begin{figure}[!htb]
 \begin{center}
   \includegraphics[scale=0.5]{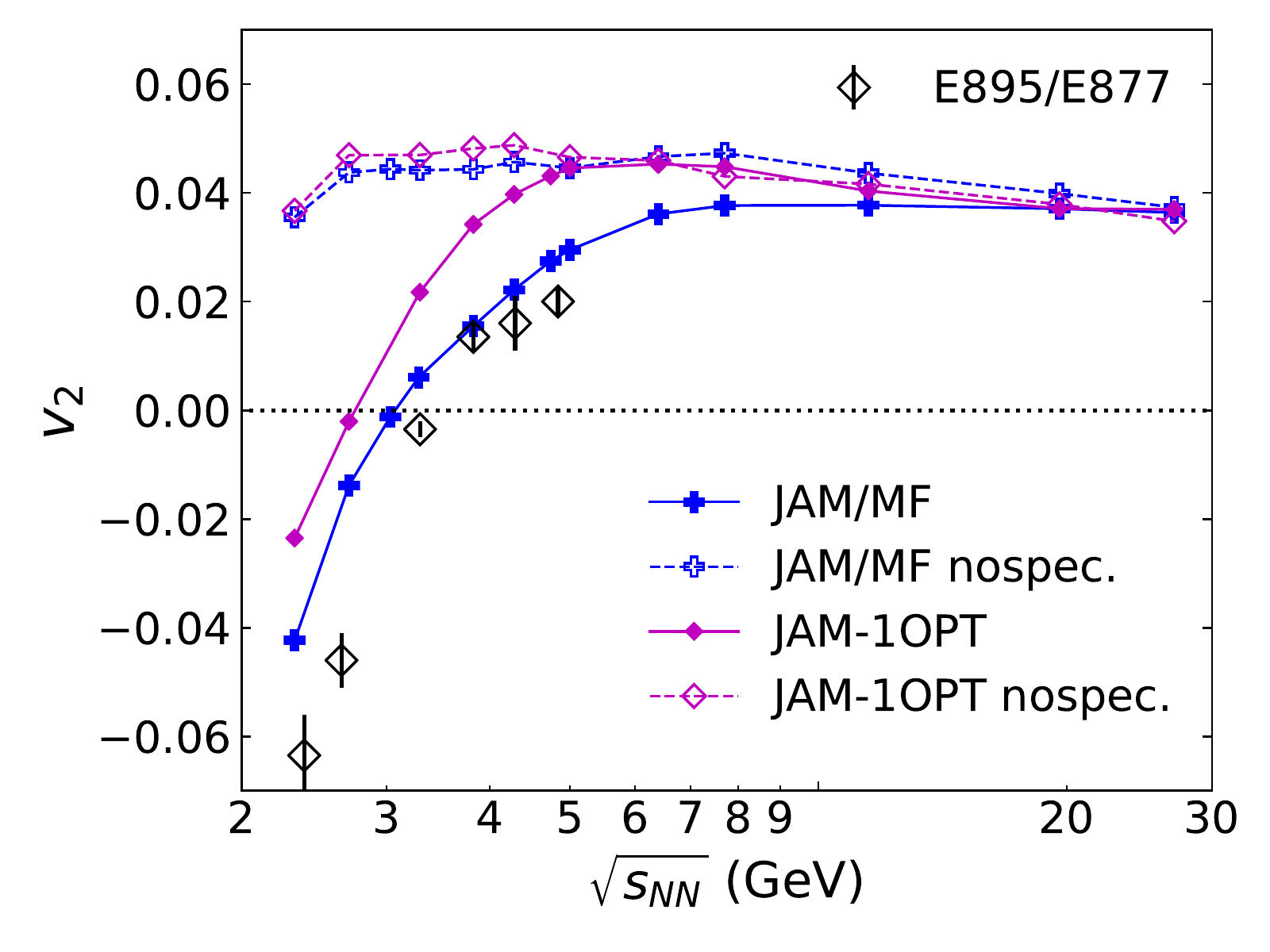}
   \caption{Beam energy dependence of $v_2$ for nucleons
   in midcentral Au+Au collisions ($4.6<b<9.4$ fm)
   from JAM first-order phase transition and mean-field simulations
   with and without spectator interactions.
   Data are take from Ref.~\cite{E895v2}.
 }
   \label{fig:v2p1m}
 \end{center}
\end{figure}

Fig.~\ref{fig:v2p1m} summarizes the effect of spectator shadowing
on the elliptic flow of nucleons at midrapidity as a function of beam energy
for mean-field simulation as well as simulations
with a first-order phase transition.
The repulsive potential generates more pressures, as a result
there is a spectator shadowing up to the beam energy of 10 GeV.
On the other hand, 
the spectator shadowing disappears due to
the strong softening effect in a first-order phase transition
around 5 GeV,
and calculations with and without spectator interaction
show almost identical results at $\sqrt{s_{NN}}> 5$ GeV.
Measurements of the elliptic flow at the beam energy range of
$5<\sqrt{s_{NN}}<10$ GeV,
which is being planed at FAIR, NICA, and J-PARC-HI,
help to extract an important information
on the pressure of the dense baryonic system created in heavy ion collisions.

\section{Summary}\label{sec:summary}

We study the role of meson-baryon and meson-meson rescattering
as well as the interaction with spectator matter on the generation of
directed and elliptic flows including EoS dependence
in Au + Au collisions at $2.3 \leq \sqrt{s_{NN}} \leq 62.4$ GeV.
It is found that initial nucleon-nucleon collisions during the passage
of two-nuclei (Glauber-type scattering)
generate negative nucleon-directed flow at the beam energy up to
27 GeV due to the spectator shadowing,
but above 27 GeV its effect becomes negligible at midrapidity.
$BM$ and $MM$ scatterings generate positive nucleon-directed flow
below 27 GeV, while they generate negative nucleon flow above 27 GeV.
The main difference in the collision dynamics below and above 27 GeV
is that rescattering happens during the passage of two nuclei below
27 GeV, while rescatterings start after passage of two nuclei
above 27 GeV (after making tilted initial condition).
Universal behavior of the nucleon directed flow at the forward
rapidity region $y>0.75y_\mathrm{c.m.}$ is observed.
Pion-directed flow is the result of the interaction between
pion and spectator nucleons,
which generate negative-directed flow for all beam energies
investigated in this work.

Our study demonstrates the importance of spectator interaction
on directed and elliptic flows
in Au+Au collisions at high baryon density region 
$\sqrt{s_{NN}}<10$ GeV for all rapidity range.
The squeeze-out effect by the spectator to the elliptic flow
becomes negligible at midrapidity at above 10 GeV.
The degree of shadowing by the spectator matter
decreases as the beam energy increases
at the forward rapidity region, and its effect becomes very small
at 62.4 GeV,
thus it supports some hydrodynamical approaches
that do not include spectators in the calculations.
Finally, we show that the enhancement of the elliptic flow
by the softening of EoS is largely due to the absence of
squeeze-out effect at $5<\sqrt{s_{NN}}<10$ GeV.
As a future study, systematic investigation of collision system
dependence is planned.

\begin{acknowledgments}

This work is supported by the MoST of China 973-Project
No.2015CB856901, NSFC under Grant No. 11575069 and 11521064.
Y. N. is supported by the Grants-in-Aid for Scientific Research
from JSPS (Nos. JP15K05079, JP15K05098, and
JP17K05448).
\end{acknowledgments}

\end{document}